\documentclass[12pt]{article}
\usepackage{amsmath,  amsthm}
\usepackage{graphicx}
\usepackage{natbib}
\usepackage{url} 
\usepackage{amssymb} 
\usepackage{color}
\usepackage{tikz}
\usepackage{caption, subcaption}
\usetikzlibrary{shapes.misc, positioning}
\newsavebox{\tempbox}

\newtheorem{proposition}{Proposition}[section]

\newcommand{\textbox}[1]
{\savebox{\tempbox}{#1}
	\ifdim\wd\tempbox<4cm\relax
	\makebox[4cm]{\usebox{\tempbox}}%
	\else
	\parbox{4cm}{\raggedright #1}%
	\fi}
\newcommand{\blind}{0}

\begin{document}
\def\spacingset#1{\renewcommand{\baselinestretch}%
{#1}\small\normalsize} \spacingset{1}

\if0\blind
{
  \title{\bf Graphical Influence Diagnostics for Changepoint Models}
  \author{Ines Wilms
  \hspace{.2cm}\\
    Department of Quantitative Economics, Maastricht University\\
    Rebecca Killick
    \\
    Department of Mathematics and Statistics, Lancaster University
    and \\
    David S. Matteson
\\
    Department of Statistics and Data Science, Cornell University}
  \maketitle
} \fi

\if1\blind
{
  \bigskip
  \bigskip
  \bigskip
  \begin{center}
    {\LARGE\bf Graphical Influence Diagnostics for Changepoint Models}
\end{center}
  \medskip
} \fi

\bigskip
\begin{abstract}
Changepoint models enjoy a wide appeal in a  variety of disciplines to model the heterogeneity of ordered data.
Graphical influence diagnostics to characterize the influence of single observations on changepoint models are, however, lacking.
We address this gap by developing a framework for investigating instabilities in changepoint segmentations and assessing the influence of single observations on various outputs of a changepoint analysis. 
We construct graphical diagnostic plots that allow practitioners to assess whether instabilities  occur; how and where they occur; and to detect influential individual observations triggering instability. 
\textcolor{black}{We analyze well-log data}
to illustrate 
how such influence diagnostic plots 
can be used in practice to reveal features of the data that may otherwise remain hidden.
\end{abstract}

\noindent%
\vfill

\newpage
\spacingset{1.5}

\section{Introduction}
\label{sec:intro}
Detecting  changes in the distributional properties of data is a common problem that arises in many application areas including; anomaly detection (e.g., \citealp{rubin2016anomaly}), bioinformatics \citep{erdman2008fast}, economics \citep{spokoiny2009multiscale}, 
genetics \citep{hocking2013learning}, network traffic analysis \citep{kwon2006wavelet}, and oceanography \citep{leeson2017}.
 The first work on such changepoint problems date back to \cite{page1954continuous}. Since then, changepoint models have been actively investigated (see \citealp{eckley2011analysis}
 for an overview) with most studies focusing on the development of changepoint detection algorithms (e.g., \textcolor{black}{the PELT method of} \citealp{killick2012optimal}  \textcolor{black}{or wild binary segmentation  of} \citealp{fryzlewicz2014}) and inference methods (e.g., \citealp{wu2020adaptive} \textcolor{black}{for univariate changepoint detection in presence of local outliers,} \citealp{grundy2020} \textcolor{black}{for a recent multivariate change in mean and variance method}). 

Recent literature is starting to consider issues that arise when applying \textcolor{black}{offline} changepoint techniques in practice such as the impact of where changepoints are incorporated into the inference pipeline; \textcolor{black}{pre-modelling as a data cleaning process, within the main modelling framework or as a post-modelling diagnostic on residuals from a fitted model} \citep{Chapman2020+}.  However, influence diagnostics--as an integral part of any data analysis--have been overlooked for \textcolor{black}{offline} changepoint analyses. Yet,  diagnostic work is vital to enable analysts and practitioners to detect
(i) potential problems with the changepoint model, 
(ii) how and where they occur and 
(iii) what may trigger these to occur. 
Providing such diagnostic tools is crucial to ensure the potential of changepoint models to be fully realized outside of the academic domain and to help practitioners
develop intuition, discover features of the data that may otherwise remain hidden and make, in the end, 
more informed decisions \citep{rajaratnam2019influence}.

In this paper, we present a unified influence framework for \textcolor{black}{offline} changepoint models that is fully aligned with this articulated need. 
The graphical influence diagnostic tools we develop are the first to highlight instabilities in changepoint models and assess the influence of single observations on the stability of the changepoint segmentation and corresponding segment parameters.
\color{black}
We devise these plots through automated 
procedures, 
available on CRAN in the \verb+R+ \citep{Rcoreteam} package \verb+changepoint.influence+,  
\color{black}
to bring the importance of influence diagnosis to the attention of researchers in the changepoint community and to stimulate their widespread usage amongst practitioners.

Model instabilities are well-known statistical problems and influence diagnostics are essential to detect them and to investigate the role of single observations that give rise to these  instabilities. 
We call observations whose alteration changes the resulting changepoint segmentation, and thus give rise to model instabilities,  \textit{influential}. 
\textcolor{black}{ Influence diagnostics have a long-standing history in regression analysis, see 
early studies by, amongst others,
\cite{cook1979influential} and \cite{belsley1980regression} 
\textcolor{black}{who assess the effect of single observations on  coefficient estimates in low-dimensional settings}
or more recent work by 
\cite{hellton2019influence},  \cite{rajaratnam2019influence} and \cite{zhao2019multiple}
\textcolor{black}{who adapt diagnostic tools to high-dimensional regression settings.}}
Developing influence diagnostics for changepoint analysis is arguably even more compelling than it is for regression analysis since influential observations can not only affect parameter estimates (for instance, segment means) but also the entire changepoint segmentation. In the same vein, several recent studies have developed influence diagnostics for variable selection procedures (e.g., \citealp{de2017detection} \textcolor{black}{for resampling-based methods} or  \citealp{rajaratnam2019influence} \textcolor{black}{for the lasso}) thereby addressing instabilities in both parameter estimation and model/variable selection.

Our contributions to the changepoint literature are twofold.
First, we introduce a new framework for diagnosing influential observations within changepoint models.
We propose two types of influence diagnostics. 
Both types \textit{alter}  individual data points and evaluate how, and to what extent, such alterations induce differences in various outputs of the changepoint analysis.
They differ in the way the data are altered.
In one extreme case, we alter data points by deleting them, one by one, thereby following the intuitive and popular deletion diagnostics \citep{belsley1980regression} for regression analysis.
In the other extreme case, we alter data points by 
\textcolor{black}{contaminating} them 
such that each point forms a segment on its own, thereby building on the idea of empirical influence functions used in robust statistics (e.g., \citealp{hampel2011robust} \textcolor{black}{for an overview} or \citealp{pison2004diagnostic} \textcolor{black}{for diagnostic plots}).  
In Section \ref{sec:framework}, we will see that \textcolor{black}{the two proposed} types of influence diagnostics provide complementary views.
Secondly, we equip researchers, analysts and practitioners working with changepoint models with a set of diagnostic plots.
These plots help
to 
visualize the output of the influence diagnostics and identify whether the original segmentation is vulnerable to instabilities. If so, in-depth plots clearly depict how and where these instabilities manifest.
More detailed follow-up visual tools then aim to identify single observations that trigger these instabilities to arise and assess their influencing role.

The remainder of this article is structured as follows.
\textcolor{black}{In} Section \ref{sec:motivatingexample}, \textcolor{black}{we} present a motivating example for the development of changepoint influence diagnostics. 
\textcolor{black}{We}  introduce the framework for diagnosing changepoint models
\textcolor{black}{in} Section \ref{sec:framework}. 
\textcolor{black}{In} Section \ref{sec:plots},  \textcolor{black}{we} present the influence diagnostic plots that guide practitioners in answering various  diagnostic questions.
\textcolor{black}{We}  demonstrate the usage of our graphical influence diagnostics on an 
\textcolor{black}{application to well-log data in Section \ref{sec:application}.}
Finally, 
\textcolor{black}{in Section \ref{sec:conclusion}, we summarize our contributions and propose several directions for
future work.}

\section{Motivating Example}\label{sec:motivatingexample}
We present a motivating example to illustrate that changepoint segmentations can be highly sensitive to individual data points, thereby calling for  appropriate influence diagnostics to identify and assess these various sources of instability and data influence on them.

\begin{figure}[t]
\captionsetup[subfigure]{labelformat=empty}
\centering
     \begin{subfigure}{\textwidth}
         \centering
         \includegraphics[width=\textwidth]{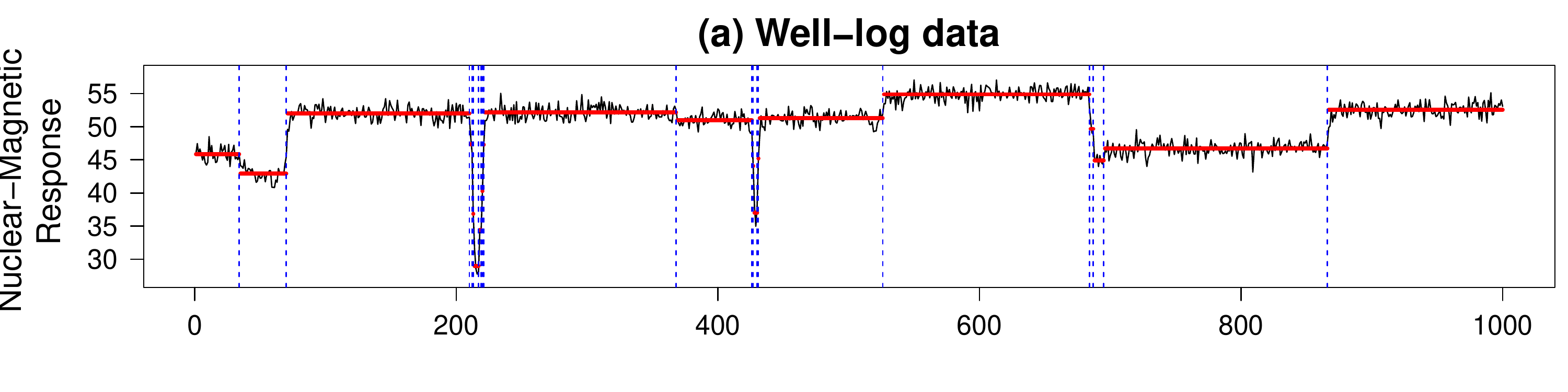}
         \caption{\label{fig:welllog}}
     \end{subfigure}
     \begin{subfigure}{\textwidth}
         \centering
         \vspace{-1.2cm}
         \includegraphics[width=\textwidth]{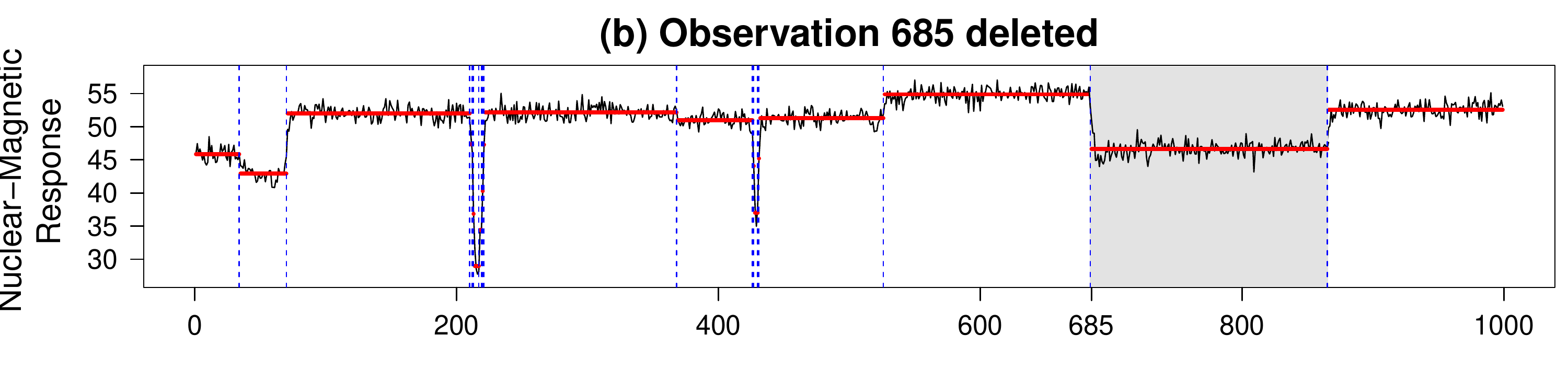}
         \caption{\label{fig:welllog_del685}}
     \end{subfigure}
          \begin{subfigure}{\textwidth}
         \centering
        \vspace{-1.2cm}
         \includegraphics[width=\textwidth]{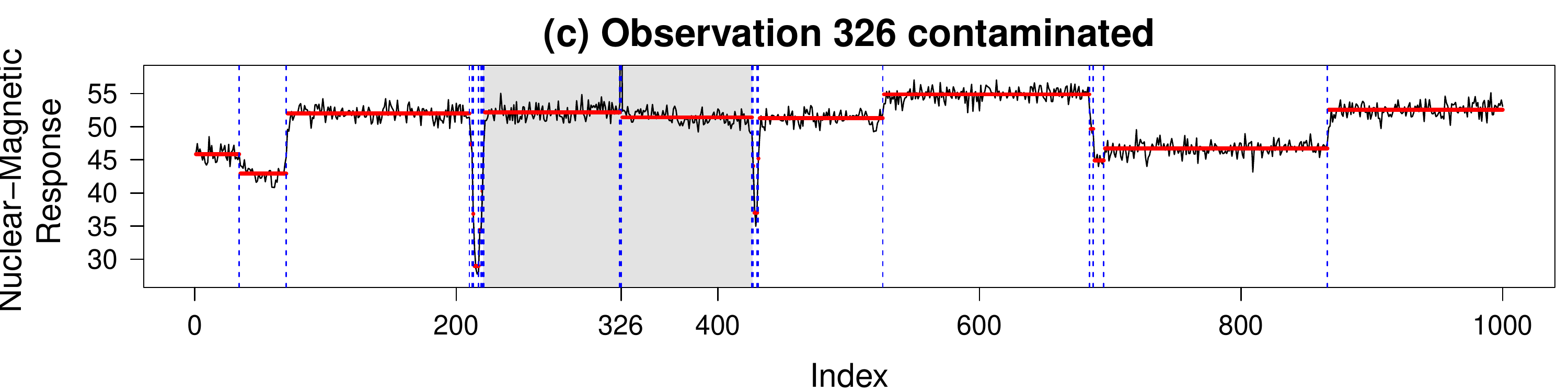}
         \caption{\label{fig:welllog_out326}}
     \end{subfigure}
     \vspace{-1cm}
\caption{(a) Well-log data with 19 changepoints (vertical dashed blue lines) and segment means (horizontal solid red lines).
(b) \textcolor{black}{Segmentation when deleting observation 685.} 
(c) \textcolor{black}{Segmentation when contaminating observation 326}. 
\textcolor{black}{The gray background in panels (b) and (c) highlights the span of changes to the segmentation  compared to panel (a).}
\label{well-log-PELT}}
\end{figure}

{\bf Well-log data.}
\textcolor{black}{We consider the problem of detecting changes in well-log data \citep{ruanaidh2012numerical}.
Figure \ref{fig:welllog} displays} 
$n=1000$ measurements from a probe that is lowered into a bore-hole. The probe takes measurements of the nuclear-magnetic response of the rock that it is passing through. Abrupt changes occur in the measurements as the probe moves from one type of rock strata to another. Changepoint analysis is used to detect these rock strata. 
\textcolor{black}{While online changepoint detection can be used to modify the settings of the drill in (near-)real time, 
we focus on influence diagnostics that are  suitable for offline changepoint analysis. The well-log data are particularly suited for this purpose as they exhibit several interesting features for influence diagnosis, as discussed below. }

{\bf Changepoint segmentation.}  
Several changepoint methods have been used to detect changes in the well-log data (\citealp{fearnhead2006exact}, and \citealp{ruanaidh2012numerical}). 
Following \cite{Fearnhead19}, we focus--throughout the paper--on the minimum penalized cost approach to detect changes in the mean and use a Normal likelihood test statistic and the Pruned Exact Linear Time (PELT) algorithm \citep{killick2012optimal}, available in
the \verb|R| 
package \verb|changepoint| \citep{killick2014changepoint}, to detect these changes.
\textcolor{black}{Details about this approach are included in Appendix A of the Supplemental Material.}
This is merely an illustrative example of a changepoint model as our framework is more broadly applicable, as will be discussed in Section \ref{sec:conclusion}.

In Figure \ref{fig:welllog}, we  visualize in \textcolor{black}{dashed (blue) lines} the  changepoints detected from a change in mean model: 
19 changes are detected. The segments vary considerably in length, ranging from segments containing as many as 171 observations (last segment) to single observation segments \textcolor{black}{(e.g., two segments in between observations 219 and 221).} 
The latter are very low measurements that occur due to  malfunctioning of the probe and can be highly influential.
\textcolor{black}{Indeed, if a data point is sufficiently extreme compared to its neighbors, it occurs}
in a segment of its own (see \citealp{Fearnhead19} \textcolor{black}{and Proposition B.1 in the Appendix).}

{\bf Changepoint stability and data influence.}
While the well-log data have been extensively analyzed through various changepoint methods, the stability of the obtained changepoint segmentation and influence of single data observations on it is less well understood.
To illustrate this, consider the influence of two types of data alterations on the obtained  segmentation. 

First, in Figure \ref{fig:welllog_del685}, we \textit{delete} data point 685. This data alteration has a drastic local impact:  17 instead of 19 changes in the mean are detected, the original changepoints at positions 687 and 695 no longer arise and all observations from position 684 to 866 are placed in a single segment instead of the original three. To reiterate, deleting a single observation removes two non-adjacent changepoints - this clearly calls into question any inference regarding those inferred changepoints and the segment means.

Second, in Figure \ref{fig:welllog_out326}, we \textcolor{black}{\textit{contaminate}}  data point 326 (in the middle of a segment) by adding twice the range of the data to its value. 
Two additional changes around the \textcolor{black}{contaminated} data point occur since ``outlying" data points are placed in their own segment - this is to be expected (\textcolor{black}{see Appendix B.2).}
Additionally and unexpectedly, the original changepoint at location 368 no longer occurs. Hence,  when observations 327-387 are not jointly considered with observations 221-326, the former are not sufficiently different from 
\textcolor{black}{the latter observations.}
Again the inference regarding the changepoint at location 368 is affected.
These two examples illustrate the dramatic impact slight data alterations or measurement errors might have on the changepoint analysis. 

{\bf Diagnostic questions.} While these motivating examples are deliberately chosen to emphasize the potential dramatic influence a \textit{single} observation can have on the output from a changepoint analysis, they raise several general 
diagnostic questions practitioners might be concerned with. We present three main diagnostic questions, each motivating the need for a particular type of influence diagnostic, as will be discussed in Section  \ref{sec:plots}:
\begin{enumerate}
\item[Q1.] \textit{Is the output of the changepoint analysis stable or  vulnerable to data instabilities?}
\item[Q2.] \textit{If vulnerable, how and where do the instabilities manifest?}
\item[Q3.] \textit{Which single influential observations trigger these instabilities to arise and how so?}
\end{enumerate}
The first two questions aim to assess the stability of various outputs of the changepoint analysis: For instance, 
which changepoints are sensitive to the data at hand and does this sensitivity  raise questions on their occurrence and/or their location? 
The third question digs deeper into the influential role of single observations on the various output measures.
An important remark that needs to be made here is that influential observations need not to be seen as harmful in the analysis, in the sense of measurement errors or extreme/atypical data points, but can be seen as data points that are highly relevant for obtaining the segmentation at hand \citep{serneels2005influence}. 

\section{Framework for Diagnosing Changepoint Models} \label{sec:framework}
In this paper, we consider observed sequences of data, $y_1,\ldots, y_n$, and assume a changepoint analysis has been performed on such a sequence resulting in $\hat{m}$ identified changepoints at ordered locations $\hat{\tau}_1,\ldots,\hat{\tau}_{\hat{m}}$.  This segmentation splits the data into $\hat{m}+1$ independent segments, the $i^{\mathrm{th}}$ of which contains the data points $y_{\tau_i +1},\ldots,y_{\tau_{i+1}}$ using the convention that $\tau_0=0$ and $\tau_{\hat{m}+1}=n$.  Our goal is to 
develop a general framework for 
assessing a segmentation's (in)stability, and
understanding the role of each observation $y_t, \ t=1, \ldots, n$,
\textcolor{black}{on the estimated segmentation.}
The framework is presented in this section. \textcolor{black}{In} Section  \ref{sec:plots}, we explain how to use the newly
constructed influence diagnostic plots for these purposes. 

\begin{figure}
\includegraphics[width=\textwidth]{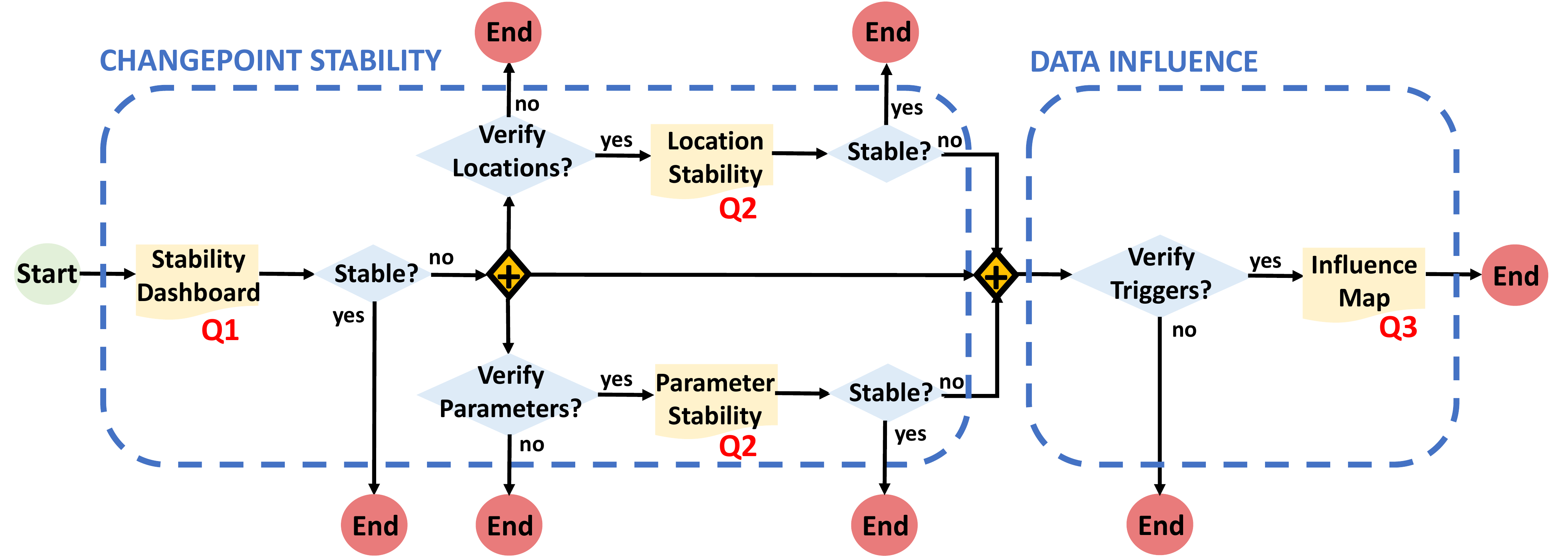}
\caption{\textcolor{black}{Workflow for assessing a segmentation's stability and identifying influence of single data points on the segmentation obtained.} \label{pic: framework}}
\end{figure}

\textcolor{black}{The proposed framework allows us to identify and analyze both global and local \linebreak (observation-specific) instabilities.}
The global diagnosis involves the assessment of the (in)stability in three pertinent outputs of a changepoint analysis: the number of changepoints, changepoint locations and segment parameters.
Monitoring changes in each of these outputs will be a useful guide towards assessing a changepoint model's (in)stability.
The more detailed diagnosis then involves the identification of single influential observations that trigger these changepoint instabilities. 

\textcolor{black}{Figure \ref{pic: framework} visualizes the typical workflow of a changepoint diagnostic analysis, thereby linking both diagnostic objectives (dashed boxes) to 
the three diagnostic questions (Q1-3 in Section \ref{sec:motivatingexample}) and the four graphical tools as indispensable workflow documents to guide practitioners in answering these questions.
The diagnostic questions and corresponding graphics are hierarchically structured from  global  towards  detailed  diagnosis.   While  the  Stability  Dashboard  is  useful  to  address  Q1,  the Location and Parameter Stability plots (additionally) address Q2, and the Influence Map is useful to (additionally) address Q3.
}

To detect changepoint (in)stabilities and quantify the effect of a single data point on them,
we follow the procedure of consecutively \textit{rolling} through all data points and at each time either \textit{deleting} 
or \textcolor{black}{\textit{contaminating}} 
a particular observation. 
This way, we 
identify 
(1) whether various output measures of the changepoint analysis are stable or differ substantially after altering individual data points;
(2) how and where  instabilities arise: locally, only affecting the segment the deleted data point is within, or global, thereby affecting other segments as well;
(3) the individual influential data points triggering these instabilities. 

\subsection{Rolling Procedure} 
For each data point $t=1, \ldots, n$ in this rolling procedure, two new segmentations are obtained.  The first new segmentation, we call the ``Observed Segmentation". Here, we simply re-run the changepoint method on the {\itshape altered} (deleted or \textcolor{black}{contaminated}) data and record the ``Observed Segmentation" obtained.
 The second segmentation, we call the ``Expected Segmentation". This segmentation corresponds to our  expectation regarding the change in a particular output measure of the changepoint analysis when either deleting a data point or 
 \textcolor{black}{contaminating} it. 
 Detailed  results for the expected segmentations under a penalized cost approach for a change in mean are provided in \textcolor{black}{Appendix B.}

While all influence diagnostic plots are constructed from the observed segmentation, only the 
\textcolor{black}{Location Stability plot and the}
Influence Map 
rely on the comparison between the  observed and expected segmentation.  
\textcolor{black}{
The idea is  that 
\textcolor{black}{these plots}
should directly highlight unusual behavior rather than changes to the segmentation we expect to see due to the data alteration. 
We therefore do not display the expected changes 
but instead compare what is observed to what is expected under a penalized cost approach such that only changes beyond the expected ones are displayed. 
As such, we draw a practitioner's attention to  insightful discrepancies that help to 
assess how and where instabilities manifest (Location Stability plot) and
quantify the influence of individual data points on a changepoint's (in)stability (Influence Map).}
The difference between both segmentations presents evidence of a single data point's influence beyond what is to be theoretically expected, which a practitioner can interpret as influence. 

\subsection{Deleting Observations}
Deletion diagnostics have been the subject of extensive research in the context of regression analysis and date back to Cook's distance \citep{cook1979influential}, 
which measures the influence of single observations on various aspects of the fitted regression model, including and excluding the observation in question. 
For changepoint models, a diagnostic analysis based on deleting data points is complicated by the fact that 
(i) individual data points can, in addition to parameter estimates as  in traditional regression analysis,  affect the entire changepoint model through the  number of changepoints and their location.
(ii) While the total number of observations $n$ might be considerable, each segment contains only a (sometimes small) fraction of the total sample size, hence individual observations can not only have a potentially tremendous local influence on the segment to which it belongs but this influence might also spill over globally to other segments. 
Hence, this calls for the need of new deletion diagnostics for detecting changepoint (in)stabilities and understanding the influential role of single data points on them. 

\textcolor{black}{Inspired by
these intuitive and popular} 
deletion diagnostics, we alter data points by deleting them, one by one, and assess the relative change in various outputs of the changepoint analysis (i.e., number of changepoints, changepoint location, segment parameters) to demonstrate the influence, or not, of the deleted data points. 
We show in \textcolor{black}{Appendix B} that the segmentation expected under a single data point deletion remains the same as the original one unless the data point belonged to a segment of length one. 
In practice, we implement the deletion approach as follows. By way of example, consider a changepoint segmentation, 1 1 1 2 3 3 3. When we leave out the first observation the expected segmentation then becomes NA 1 1 2 3 3 3.  
However, when we leave out the fourth observation, the expected segmentation is 1 1 1 NA 2 2 2. 
Note that we re-number the segments to ensure that two neighbouring segments differ in their numbering by one.

\subsection{\textcolor{black}{Contaminating} Observations}
As an alternative to deletion diagnostics, empirical influence functions are commonly used in 
robust statistics to determine, on a sample-specific basis, the influence of each data point on parameter estimation or prediction \citep{hampel2011robust}. 
To this end,  the effect of an infinitesimal contamination at a certain data point on a  statistical functional of interest is measured and used as a diagnostic tool to assess its influence. 
It is hereby crucial to stress the discrepancy between influence and extremeness. While both properties  coincide in the detection of influential outliers (i.e., atypical 
data points), in general, non-outlying influential data points as well as non-influential outliers do exist (\citealp{serneels2005influence}). 

\textcolor{black}{Inspired by techniques from}
robust statistics, we alter data points 
one by one such that each point is made atypical/outlying and assess the relative change in various changepoint outputs of this \textcolor{black}{contaminated} data point. 
\cite{fearnhead2006exact} showed that the segmentation expected under this data alteration corresponds to the segmentation obtained on the original data with two extra changes  added before and at the \textcolor{black}{contaminated} position. Being different from the bulk of the data, a \textcolor{black}{contaminated} point thus warrants its own segment. However, only one extra change occurs if we are close to an original changepoint.
Coming back to our earlier example (i.e., segmentation 1 1 1 2 3 3 3), when we 
\textcolor{black}{contaminate} the first observation the expected segmentation  becomes 1 2 2 3 4 4 4, hence there are  four segments in total instead of three.  
However, when we \textcolor{black}{contaminate} the second observation, the expected segmentation is 1 2 3 4 5 5 5, thereby including two additional changes.

\smallskip
These two ways of altering observations provide two extreme perspectives:
on the one hand, when deleted, 
the segmentation reveals what would have happened had the observation not been observed.  On the other hand, when \textcolor{black}{contaminated}, the point is simultaneously both maximally influential (it has its own segment) and minimally influential (in its own segment it does not  directly contribute to other segments).  Curiously the \textcolor{black}{contaminated} point also forces a shortening of the segments on either side, thus allowing one to identify if a change is sensitive to the length of the segment it is within.  
Thereby, both present complementary views on the stability of changepoint segmentations 
and allow us to better grasp the overall influence of single observations.

\section{Influence Diagnostic Plots} \label{sec:plots}
We create a \textit{set of four diagnostic plots} which range from coarse level to detailed, namely the ``Stability Dashboard", ``Segment Location Stability" and  ``Segment Parameter Stability" plots, and finally the ``Influence Map". 
The different plots each aim to tackle a specific diagnostic question (see Figure \ref{pic: framework}), making the choice of an appropriate plot crucial for highlighting a particular aspect of the influence diagnosis for changepoint models.
Practitioners should choose the most appropriate level of detail for the data set and question they are considering. 
All plots rely on the rolling procedure discussed in Section \ref{sec:framework} and are constructed for both the case where data points are consecutively \textit{deleted} and \textit{\textcolor{black}{contaminated}}.
The influence diagnostic plots can be created via the R package \texttt{changepoint.influence}.

\begin{figure}[t]
\centering
\includegraphics[width=\textwidth]{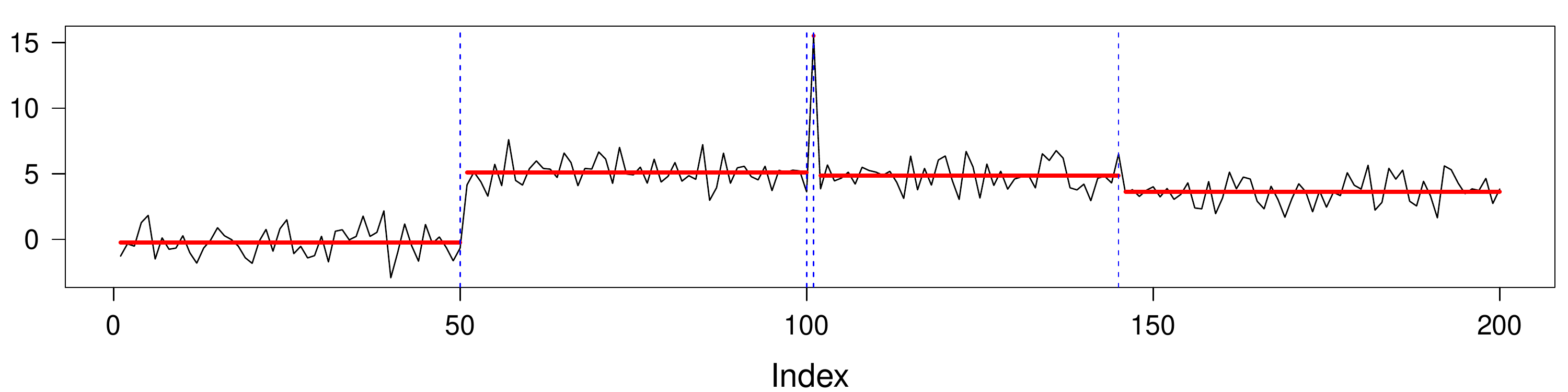}
\caption{Simulated data example with four changepoints (vertical dashed lines). The horizontal solid lines are the segment means. 
\label{simulated_data_plot}}
\end{figure}

To illustrate the usage of the plots and provide guidance on how to interpret their various features, we make use of a simulated data example. We generate an ordered sequence of length $n=200$ with four changes in mean: the first 50 data points are generated from a standard Normal distribution. Data points 51 to 100 as well as data points 102 to 150 are drawn from a Normal distribution with mean five and unit variance. 
An atypical data point, drawn from a Normal distribution 
with \textcolor{black}{mean 15 (10 standard deviations above the mean either side)} 
is included at 101. Finally, the last 50 data points are drawn from a Normal distribution with mean four and unit variance, giving rise to a relatively small change in mean.
The simulated data sequence is shown in Figure \ref{simulated_data_plot} together with the four changepoints (at positions 50, 100, 101, 145) detected by the Normal likelihood test statistic with the PELT search method.

\subsection{Stability Dashboard}
At the coarsest level we have the
``\textit{Stability Dashboard}": it presents the original data with the original changepoints depicted as vertical lines at the changepoint locations. 
\textcolor{black}{The lines are now displayed either dashed (green), dot-dashed (orange) or dotted (red),}
in line with the extent to which each is vulnerable to data instabilities. 
We use \textcolor{black}{dashed} green for unaffected changepoints (\textcolor{black}{i.e., unaffected when all  data points other than itself are altered}), \textcolor{black}{dot-dashed} orange when a changepoint moves or is deleted (for at least one altered point) and \textcolor{black}{dotted} red when a changepoint forms a segment on its own. 
This plot thus provides a coarse depiction of the results to directly address the first diagnostic question. 

\begin{figure}
\centering
\begin{subfigure}[b]{0.49\textwidth}
\includegraphics[width=\textwidth, page = 1]{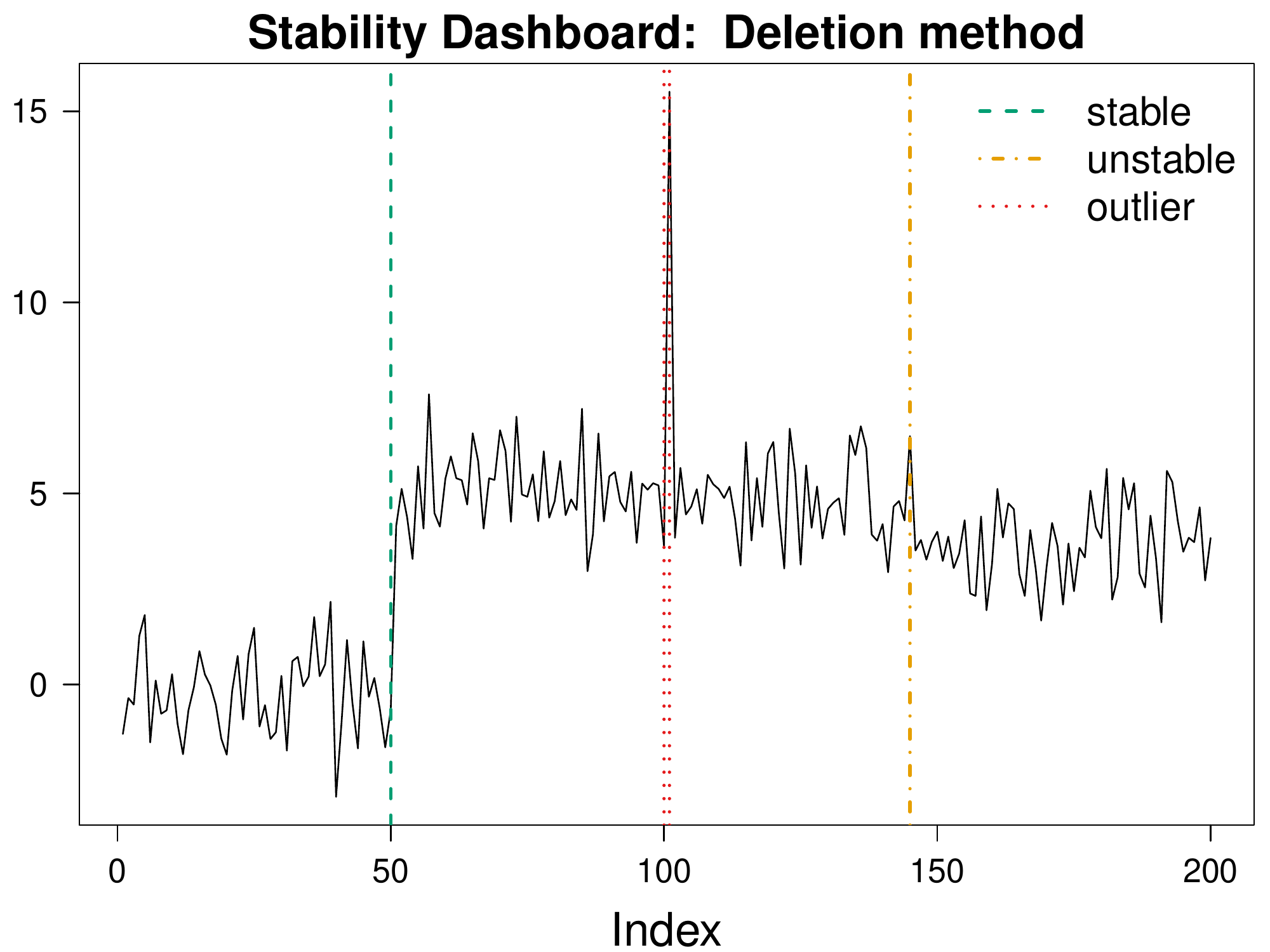}
\caption{\label{fig:sim30:stabdash:del}}
\end{subfigure}
\begin{subfigure}[b]{0.49\textwidth}
\includegraphics[width=\textwidth, page = 2]{Figure4.pdf}
\caption{\label{fig:sim30:stabdash:out}}
\end{subfigure}
\caption{Stability Dashboard  when (a) deleting and (b) \textcolor{black}{contaminating} observations. 
\label{fig:stabdash}}
\end{figure}

The Stability Dashboards for the simulated data example are presented in Figure \ref{fig:stabdash}. 
Both the deletion and outlier method give the same insights (though deviate as we delve deeper): the first changepoint is stable (dash green), the last is somewhat unstable (dot-dash orange), and changepoints at positions 100 and 101 are \textcolor{black}{bounding an outlier (dot red)} Whether a practitioner continues or not with inspecting the more detailed diagnostic plots depends on the outcome of the Stability Dashboard. If all the changepoints are stable (green) then the segmentation seems stable and there may be no need to delve further into the results.  
In contrast, if some  changepoints are
\textcolor{black}{unstable or outlying} then the practitioner may wish to further investigate  how and where these instabilities manifest. 

\subsection{Segment Location Stability}
We next move to the second level to assess how and where instabilities in the location of the changepoints occur.
To this end, 
\textcolor{black}{Location Stability plot of} the changepoint locations across the $n$ altered data points can be used.  

\textcolor{black}{We record the} number of times a changepoint alteration occurs as well as the location of any moved or additional changepoints.  This plot thus allows practitioners to assess how and where instabilities manifest, thereby (partially) addressing the second diagnostic question. 

For a segmentation which does not vary across altered data points, we  expect \textcolor{black}{each of the original changepoints to stay in place when all data points other than itself are altered.
For ease of use, we directly display the discrepancy in number of changepoint occurrences from this expected maximum.
This way only unstable (dot-dashed orange) or outlying (dotted red) original changepoints may enter the plot with a negative difference. The latter indicates that the changepoint no longer occurs at its original location for some instances. Either the changepoint moves to another location or it disappears completely.
When a changepoint moves, it will be offset and depicted by a positive difference at another location, thereby leading towards a net balance of zero. Disappearing changepoints, on the other hand, do not appear in the plot but can be deduced from net negative balances. It can also occur that the original changepoints remain but that additional changepoints occur due to the alterations (net positive) but this is much less common in our experience.}

\begin{figure}[t]
\centering
\begin{subfigure}[b]{0.49\textwidth}
\includegraphics[width=\textwidth, page = 1]{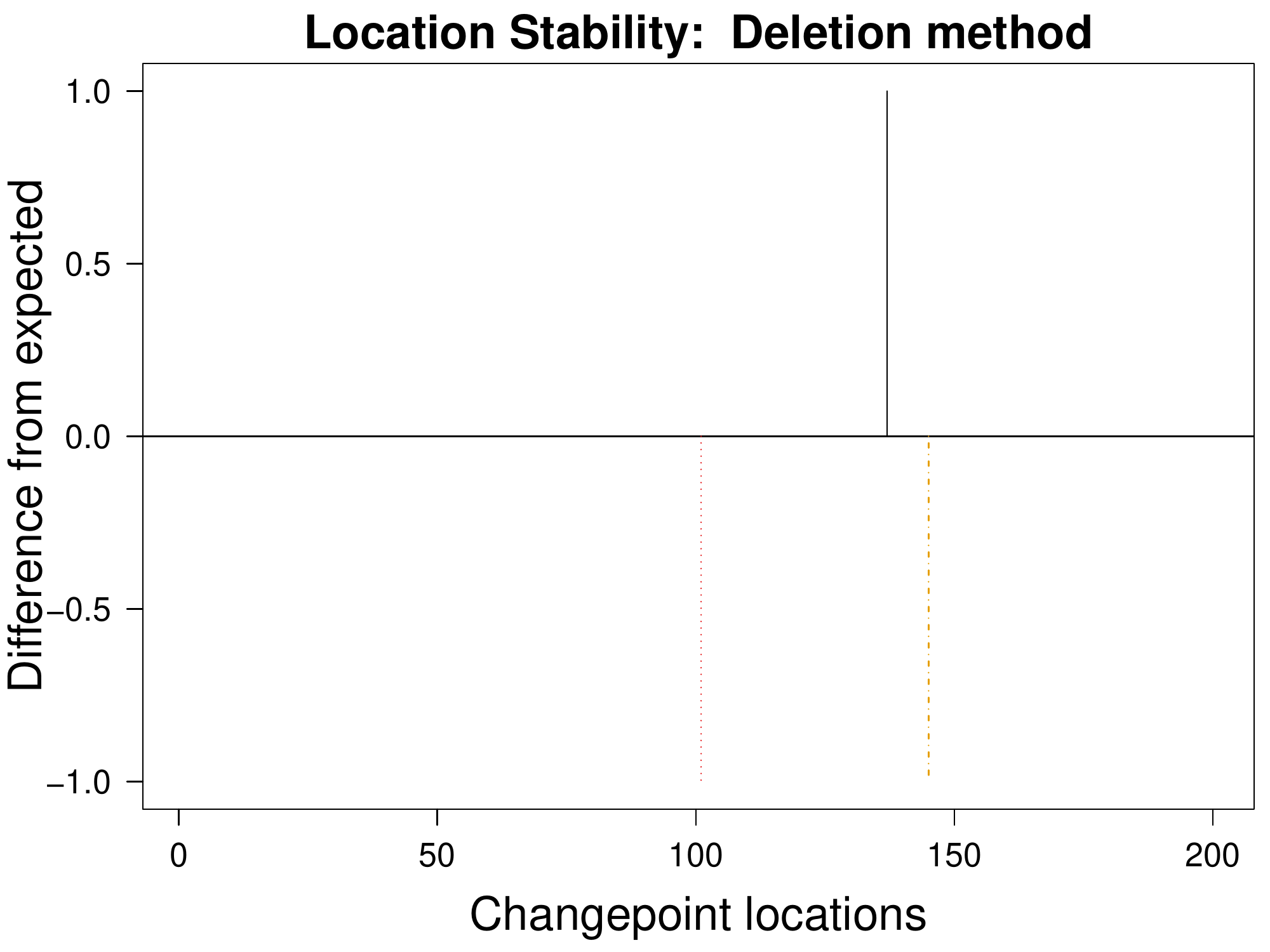}
\caption{\label{fig:sim30:stabloc:del}}
\end{subfigure}
\begin{subfigure}[b]{0.49\textwidth}
\includegraphics[width=\textwidth, page = 2]{Figure5.pdf}
\caption{\label{fig:sim30:stabloc:out}}
\end{subfigure}
\caption{Location Stability plot of the simulated data when (a) deleting and (b) \textcolor{black}{contaminating} observations (right).
\label{fig:sim30:stabloc}}
\end{figure}

The Location Stability plots for the simulated data example are presented in Figure \ref{fig:sim30:stabloc}. 
When deleting observations one by one, changepoint 145 turns out to be slightly unstable as it moves earlier for one instance, \textcolor{black}{as can be seen from 
the negative dot-dashed (orange) line of height one at the changepoint location which is offset by the positive solid (black) line of the same height at location 137, see Figure \ref{fig:sim30:stabloc:del}.}
When \textcolor{black}{contaminating} observations one by one, changepoint 145 shows more instability as it 
\textcolor{black}{displays differences from what is expected on more than 25 instances. On six instances, the changepoint is moved earlier, as can be seen from the  black positive line of height six in Figure \ref{fig:sim30:stabloc:out}.
 The net negative balance indicates that}
 the changepoint disappeared completely for several data \textcolor{black}{contaminations}.  The negative red line of length one at location 101 (Figure \ref{fig:sim30:stabloc:del})  indicates that the changepoint induced by the outlying data point at this location disappears completely when it is deleted.
  These plots give an indication of {\itshape what} the stability looks like but not \textcolor{black}{{\itshape which} data points influence this behaviour}.

\subsection{Segment Parameter Stability}
This plot complements the previous in tackling the second diagnostic question by considering  instabilities in the segment parameters, such as the mean. It is important to investigate the segment parameters separately as the changepoint locations may vary but for a small or uncertain changepoint, the segment parameters may not vary considerably.  If one is only interested in inference on the segment parameters and not the changepoint locations then this is important information.

To construct our diagnostic plot 
we start by depicting the original segment parameters, such as the mean in our example, by \textcolor{black}{solid (red) lines,} 
which correspond to the ones from Figure \ref{simulated_data_plot}. 
On the same plot we add the mean of each data point across the $n$ 
data alterations over time.  For many segmentations where the altered point is far away, there will be no difference in the mean values.  We thus take the unique values of the segment means and plot them in shades of gray, scaled by the frequency of occurrence.  Thus common values across many iterations of the rolling procedure appear darker than those across few iterations. \textcolor{black}{The original parameter estimate is intentionally thick on the plot to ensure that any black seen around this is meaningful.} The wider the dark area (vertically) that can be seen around the original segment means (red), the more evidence of instability.  

\begin{figure}
\centering
\begin{subfigure}[b]{0.49\textwidth}
\includegraphics[width = \textwidth, page = 1]{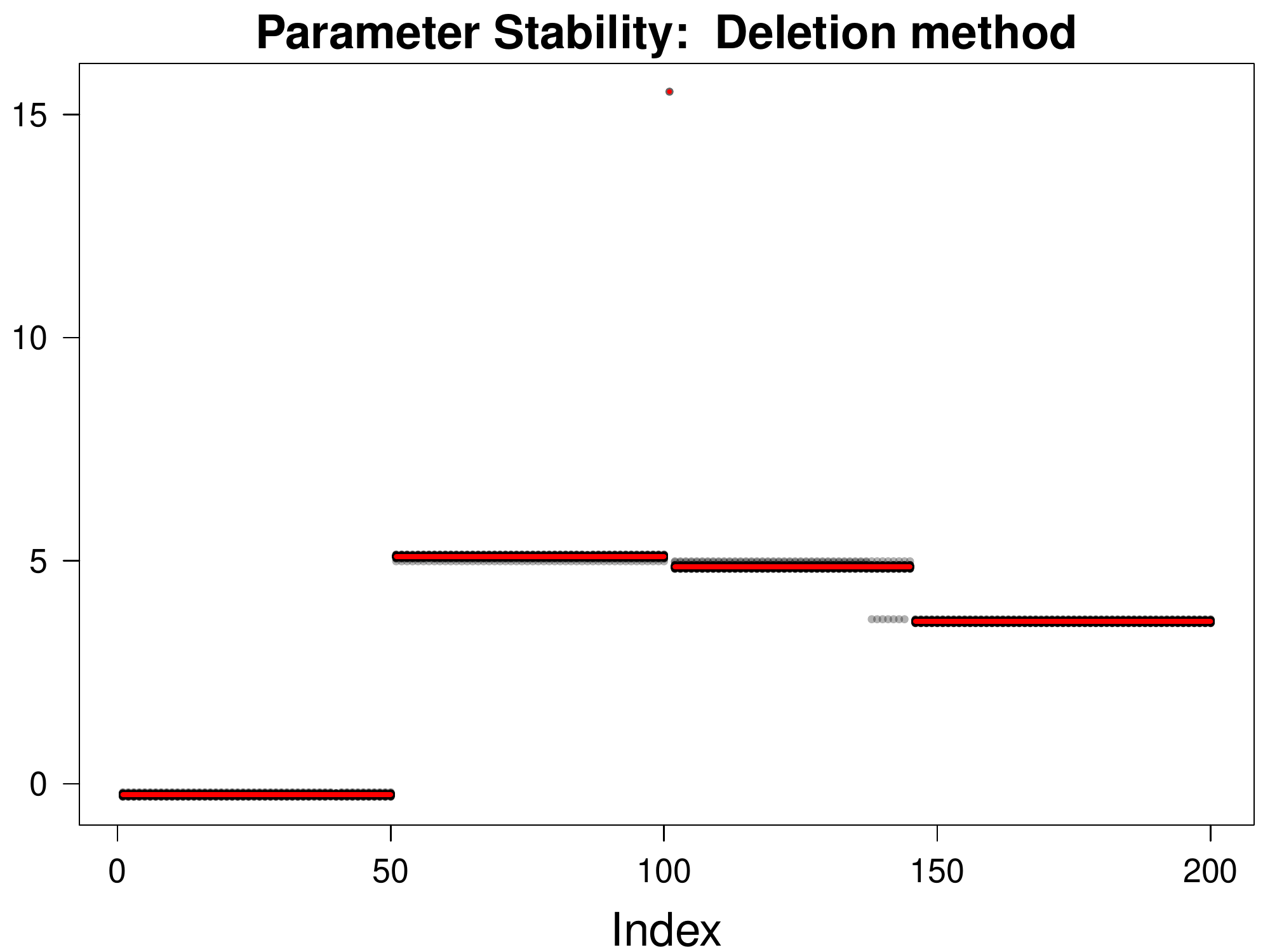}
\caption{\label{fig:sim30:stabparam:del}}
\end{subfigure}
\begin{subfigure}[b]{0.49\textwidth}
\includegraphics[width = \textwidth,  page = 2]{Figure6.pdf}
\caption{\label{fig:sim30:stabparam:out}}
\end{subfigure}
\caption{Parameter Stability plot of the simulated data when (a) deleting and (b) \textcolor{black}{contaminating} observations.  \label{fig:sim30:stabparam}}
\end{figure}

The Parameter Stability plots for the simulated data example are presented in Figure \ref{fig:sim30:stabparam}. 
For the deletion method (Figure \ref{fig:sim30:stabparam:del}), the means across the data deletions appear very tightly around the original means, thereby supporting stability of the segment means. Only for data points 137-144, a minor instability is observed by the additional dark (lower) line which occurs at the height of the next segment's mean and is caused by changepoint 145 being moved earlier.

For the \textcolor{black}{outlier} method (Figure \ref{fig:sim30:stabparam:out}), the darker areas are typically larger in size, especially towards the edges of a segment. This is in line with our expectation, since the data contamination induces two additional changes thereby triggering directly surrounding segments to be smaller in size. The additional variability then arises due to the fewer data points available for estimation of the segment parameters. \textcolor{black}{This results in a ``bleeding" effect at the edges of the segments.} The most pronounced instability in segment means is again observed for the data points around the last changepoint.  Coupled with the stability dashboard, it is clear that the changepoint moves both earlier (producing a lower mean) and disappears (producing a higher mean after 145) although again it gives no information as to {\itshape which} observations are responsible for this.

\subsection{Influence Map} 
At the most detailed level we have an ``\textit{Influence Map}" of 
salient differences between the observed and expected segmentation from the deleted or \textcolor{black}{contaminated} data points. This final plot \textcolor{black}{is a heat map which} identifies single, influential observations that trigger changepoint instabilities, thereby addressing the third diagnostic question.
The \textcolor{black}{heat map}  depicts the difference in segment number between the observed and expected segmentations across each of the altered data points.  
Analogous influence maps can be made for other outputs of the changepoint analysis. 

The horizontal axis of the Influence Map is the standard time index of the original data (1 to $n$).  The vertical-axis indexes the altered data point (1 to $n$). 
\textcolor{black}{Each coloured (taupe or blue) pixel} marks the difference between the observed and expected segmentation at the specific $(x,y)$ co-ordinate.  
The data point on the vertical-axis should be understood as the influential data point whose alteration leads to changes in the affected data points on the horizontal axis if any colouring appears.
We colour zero difference \textcolor{black}{in the heat map} as white, increases in segment number as \textcolor{black}{taupe} and decreases  as blue.  
Hence, data points on the vertical axis without a single coloured co-ordinate on the horizontal axis can be considered as non-influential since they do not trigger any changepoint instability. 
Rows with coloured 
\textcolor{black}{pixels} correspond to data points which are instability triggers. 
The intensity of the colour then signifies the magnitude of the instability (namely the increase or decrease in segment number).
Coloured areas are expected to occur around unstable or outlying (orange or red) changepoints, which are depicted as coloured circles on the diagonal. 

\begin{figure}[t]
\centering
\begin{subfigure}{0.43\textwidth}
\includegraphics[width=\textwidth, page = 1]{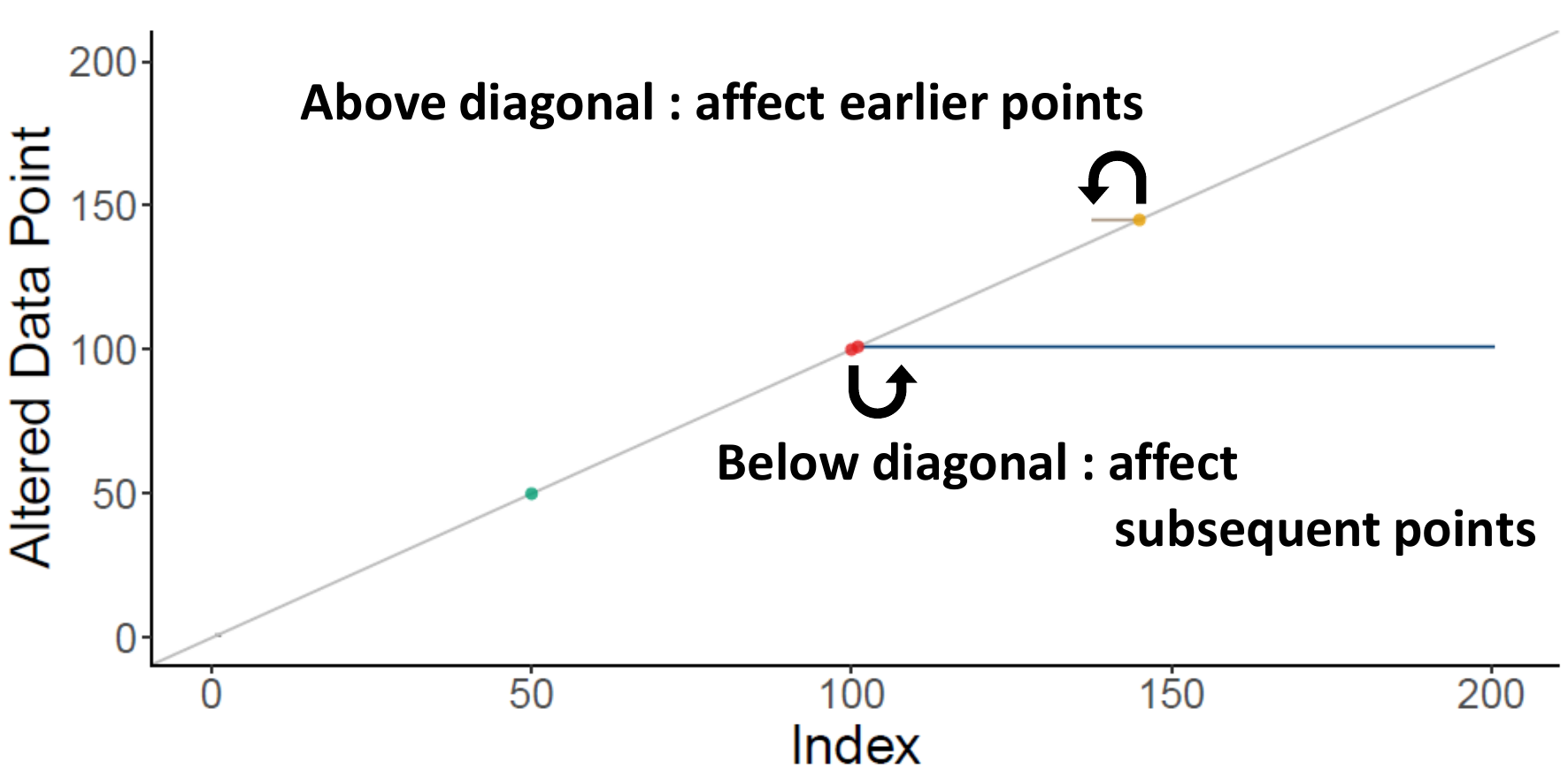}
 \caption{\label{fig:infl_features1}}
\end{subfigure}
\hspace{0.5cm}
\begin{subfigure}{0.43\textwidth}
\includegraphics[width=\textwidth, page = 2]{Figure7.pdf}
 \caption{\label{fig:infl_features2}}
\end{subfigure}

\begin{subfigure}{0.43\textwidth}
\includegraphics[width=\textwidth, page = 4]{Figure7.pdf}
 \caption{\label{fig:infl_features3}}
\end{subfigure}
\hspace{0.5cm}
\begin{subfigure}{0.43\textwidth}
\includegraphics[width=\textwidth, page = 3]{Figure7.pdf}
 \caption{\label{fig:infl_features4}}
\end{subfigure}
\caption{Main Features of the Influence Map.} \label{features_influence_map}
\end{figure}

Before discussing the Influence Maps for the simulated data example, we describe its important features to aid practitioners in studying the influential role of the individual data points through these maps. These features are summarized in Figure \ref{features_influence_map}.
(i) Figure \ref{fig:infl_features1} highlights the role of the \textit{diagonal}:
colouring above the diagonal indicates that an alteration of the corresponding data point (on the vertical axis) affects earlier data points, colouring below the diagonal indicates that subsequent data points are affected.
(ii) Figure \ref{fig:infl_features2}  concerns the \textit{horizontal span} of the colouring: 
\textcolor{black}{a  stop in colouring} indicates that changepoints have moved, while a  continuation \textcolor{black}{of colouring} to the last data point indicates that, in total, fewer or additional changepoints are detected.
(iii) Figure \ref{fig:infl_features3} zooms in on the discrepancy between \textit{local versus global} effects.
Most colouring originates on the diagonal, thereby indicating that a data point's alteration mainly affects neighbouring data points that most often belong to the same segment. 
By contrast, in some cases a coloured 
\textcolor{black}{pixel} may originate away from the diagonal, thereby exercising global influence. 
(iv) Finally, Figure \ref{fig:infl_features4} zooms in on the \textit{height} of \textcolor{black}{the colouring}.
All data points (on the vertical axis) that
\textcolor{black}{appear in the } coloured area are influential and assert influence over the corresponding data points on the horizontal axis. The height can be seen as the extent to which instability arises in this influential region.

\begin{figure}[t]
\centering
\begin{subfigure}{0.49\textwidth}
\includegraphics[width=\textwidth, page = 1]{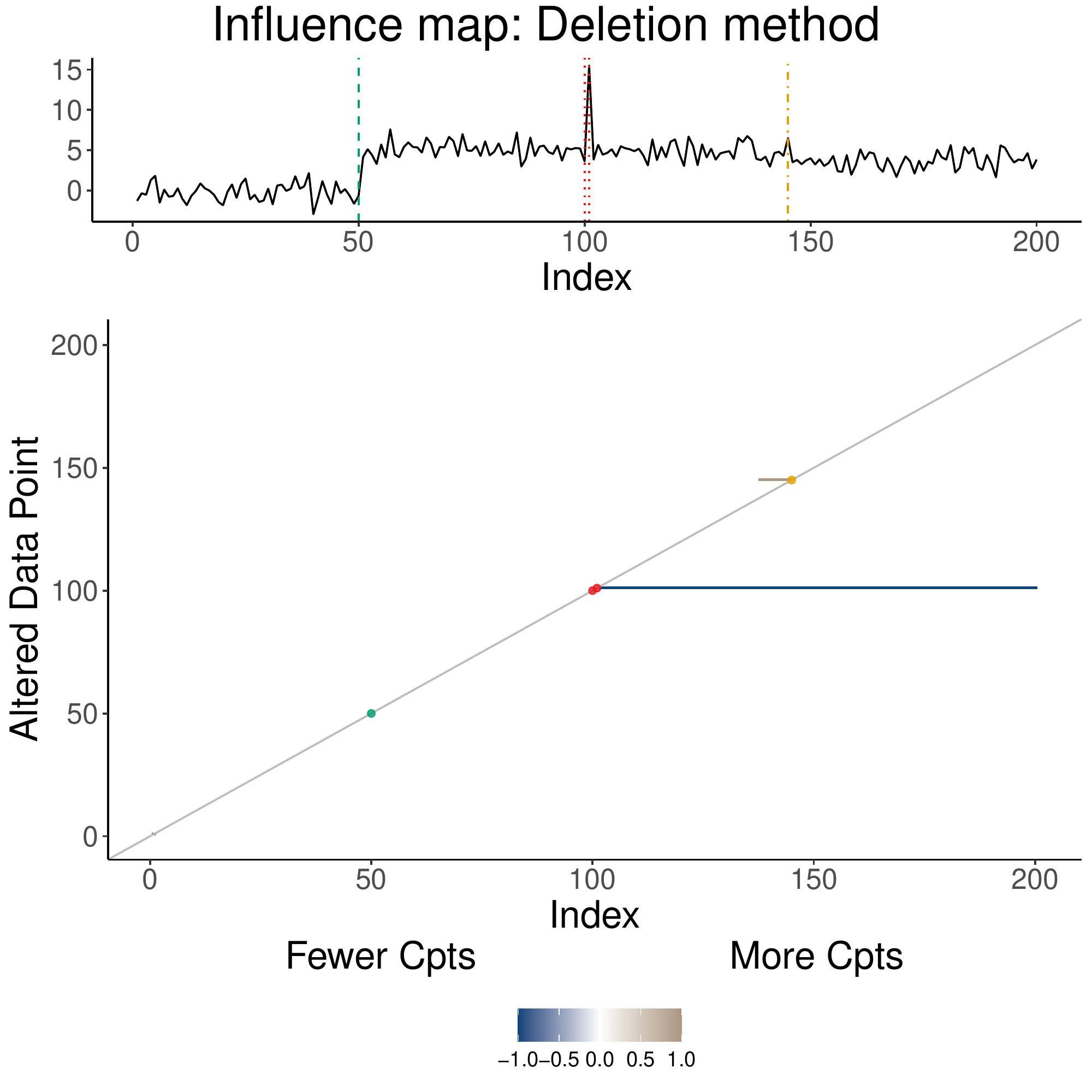}
 \caption{\label{fig:sim30:inflmap_del}}
\end{subfigure}
\begin{subfigure}{0.49\textwidth}
\includegraphics[width=\textwidth, page = 2]{Figure8.pdf}
 \caption{\label{fig:sim30:inflmap_out}}
\end{subfigure}
\caption{Influence Map of the simulated data when (a) deleting  and (b)  \textcolor{black}{contaminating} observations. 
\label{simulated_data_influence_map}}
\end{figure}

Relying on these features, we are ready to discuss the Influence Maps for the simulated data example, as presented in Figure \ref{simulated_data_influence_map}. 
Across both maps,  few coloured areas appear, each of them characterizing some form of instability. 
All of them occur around the originally detected unstable or outlying changepoints. 
The instability triggers (i.e., data points on vertical axis with colouring) are observations 101 and 129-157; their influential role will be detailed below.
Note that most coloured areas are blue, thereby indicating that a particular data point (on the horizontal axis) has a lower segment number in the observed segmentation than expected; in other words less than expected changepoints occur.  We subsequently discuss the Influence Maps according to their main features.

(i) \textcolor{black}{In this example, we see that influential data points have a tendency to affect subsequent data points rather than preceding ones since  most colouring occurs below the diagonal.}
Consider the 
\textcolor{black}{blue colouring in Figure \ref{fig:sim30:inflmap_del}.}
When deleting outlying data point 101 (i.e., the instability trigger on the vertical axis), the changepoint induced by it disappears, and thus all subsequent observations (horizontal axis) are affected by having a lower segment number than expected. The Influence Map highlights this data instability \textcolor{black}{through the blue colouring of all pixels until the last data point}
and is in line with 
\textcolor{black}{the negative dotted red line of height one in the Location Stability plot (Figure \ref{fig:sim30:stabloc:del}).}

(ii) We find evidence of influential data points moving original changepoints as well as triggering some to disappear. 
To this end, we zoom in on the instability of changepoint 145.
When deleting this data point, 
Figure \ref{fig:sim30:inflmap_del} (\textcolor{black}{taupe pixels} above diagonal line) shows that the changepoint gets moved earlier towards position 137; 
\textcolor{black}{in line with the black positive height in Figure \ref{fig:sim30:stabloc:del}.}
The same change in changepoint location occurs when \textcolor{black}{contaminating} observations 129 to 135, as can be seen from the \textcolor{black}{taupe} below-diagonal \textcolor{black}{colouring} in Figure \ref{fig:sim30:inflmap_out}. 
By contrast, the changepoint disappears when observations 136 to 157 (vertical axis) are \textcolor{black}{contaminated} (blue zone).

(iii) Almost all colouring originates on the diagonal, thereby indicating that a data point's alteration mainly affects its neighbors of the same segment. An exception is the \textcolor{black}{contamination} of observations 129 to 135 (right panel) which makes observations 138 to 144 move from its original fourth segment to the fifth segment. 

(iv) Finally, two influential regions appear in Figure \ref{fig:sim30:inflmap_out}. The blue influential region is the most outspoken one:  the \textcolor{black}{contamination} of no less than 21 data points (i.e., instability triggers 136-157 on vertical axis) all trigger changepoint 145 to disappear; 
\textcolor{black}{in line with the net negative balance for changepoint 145 in Figure \ref{fig:sim30:stabloc:out}.}
The smaller red region indicates that the \textcolor{black}{contamination} of six data points (i.e., instability triggers 129-135 on vertical axis) causes the changepoint at location 145 to move earlier; 
\textcolor{black}{in line with the black positive difference of} height six at position 137 in the Location Stability plot (Figure \ref{fig:sim30:stabloc:out}).

\section{Well-log Application} \label{sec:application}
We now return to the well-log data, presented in Section \ref{sec:motivatingexample}, and address our main diagnostic questions one by one.

\subsection{Stability of the Changepoint Analysis}
\begin{figure}
\centering
\begin{subfigure}{0.49\textwidth}
\includegraphics[width=\textwidth, page = 1]{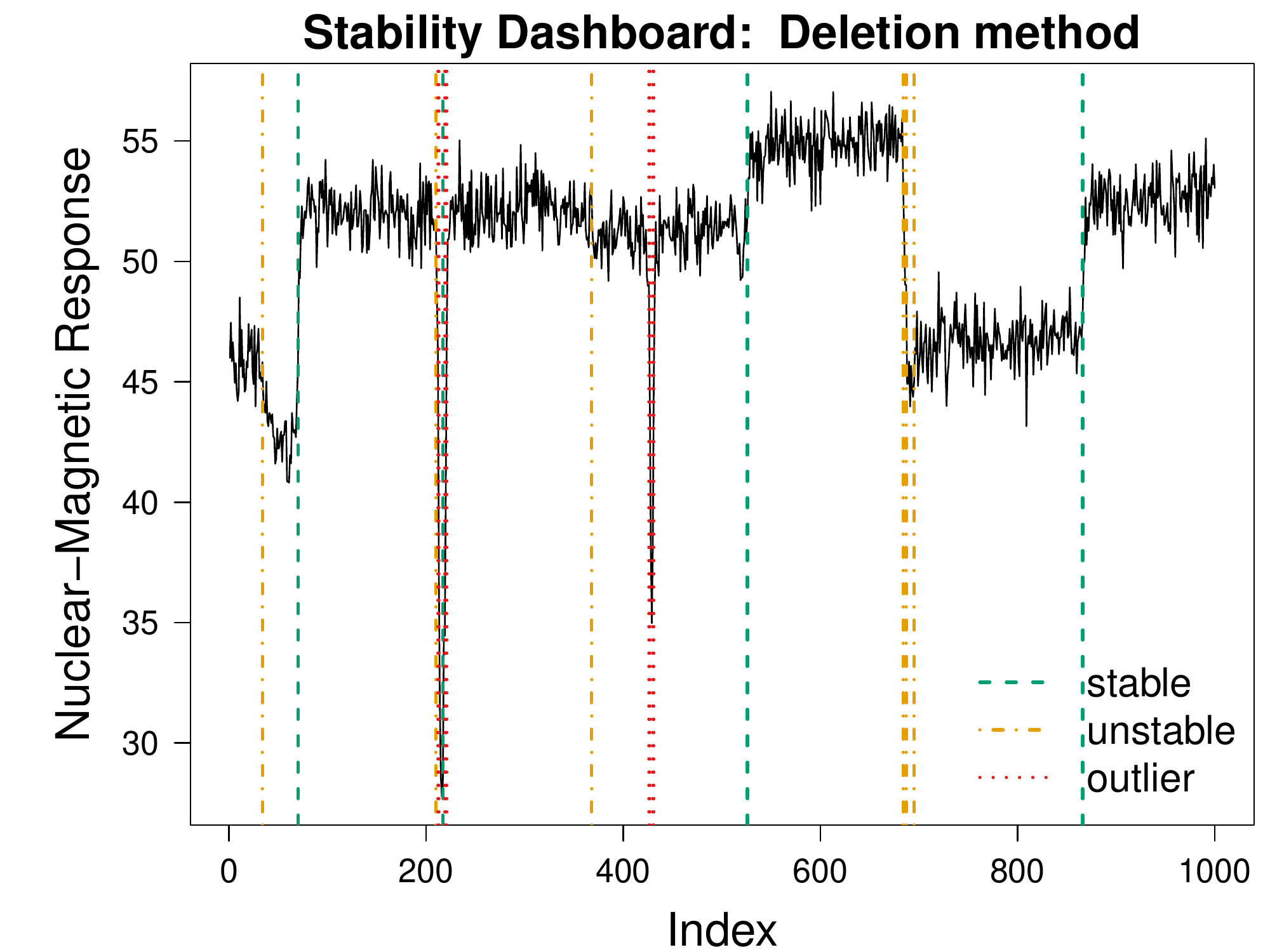}
 \caption{\label{fig:welllog:stabdash:del}}
\end{subfigure}
\begin{subfigure}{0.49\textwidth}
\includegraphics[width=\textwidth, page = 2]{Figure9.pdf}
 \caption{\label{fig:welllog:stabdash:out}}
\end{subfigure}
\caption{\label{WellLogStabDash}Stability Dashboards for the Well-log data.}
\end{figure}

We start by tackling our general diagnostic question ``\textit{Is the output of changepoint analysis stable or vulnerable to data instabilities?}" through the Stability Dashboards, presented in Figure \ref{WellLogStabDash}. Out of the original 19 changepoints, 15 are depicted as potentially unstable (orange or red) by the deletion method (Figure \ref{fig:welllog:stabdash:del}) while all but one (change at location 217) are coloured orange or red by the \textcolor{black}{outlier} method (Figure \ref{fig:welllog:stabdash:out}).

While such pronounced results will not arise for each changepoint analysis, they do illustrate that influence diagnostics should not be overlooked but rather considered as a much needed natural successor to any changepoint analysis. The mere visualization of one single additional graphic, the Stability Dashboard, can either re-assure  practitioners on the stability of their performed analysis or warn them for the occurrence of instabilities. 
In the latter case, a  more detailed influence diagnosis can be performed through our other diagnostic tools which are discussed next. 

\subsection{Manifestation of the Instabilities}

\begin{figure}[t]
\centering
\begin{subfigure}[b]{0.49\textwidth}
\includegraphics[width=\textwidth, page = 1]{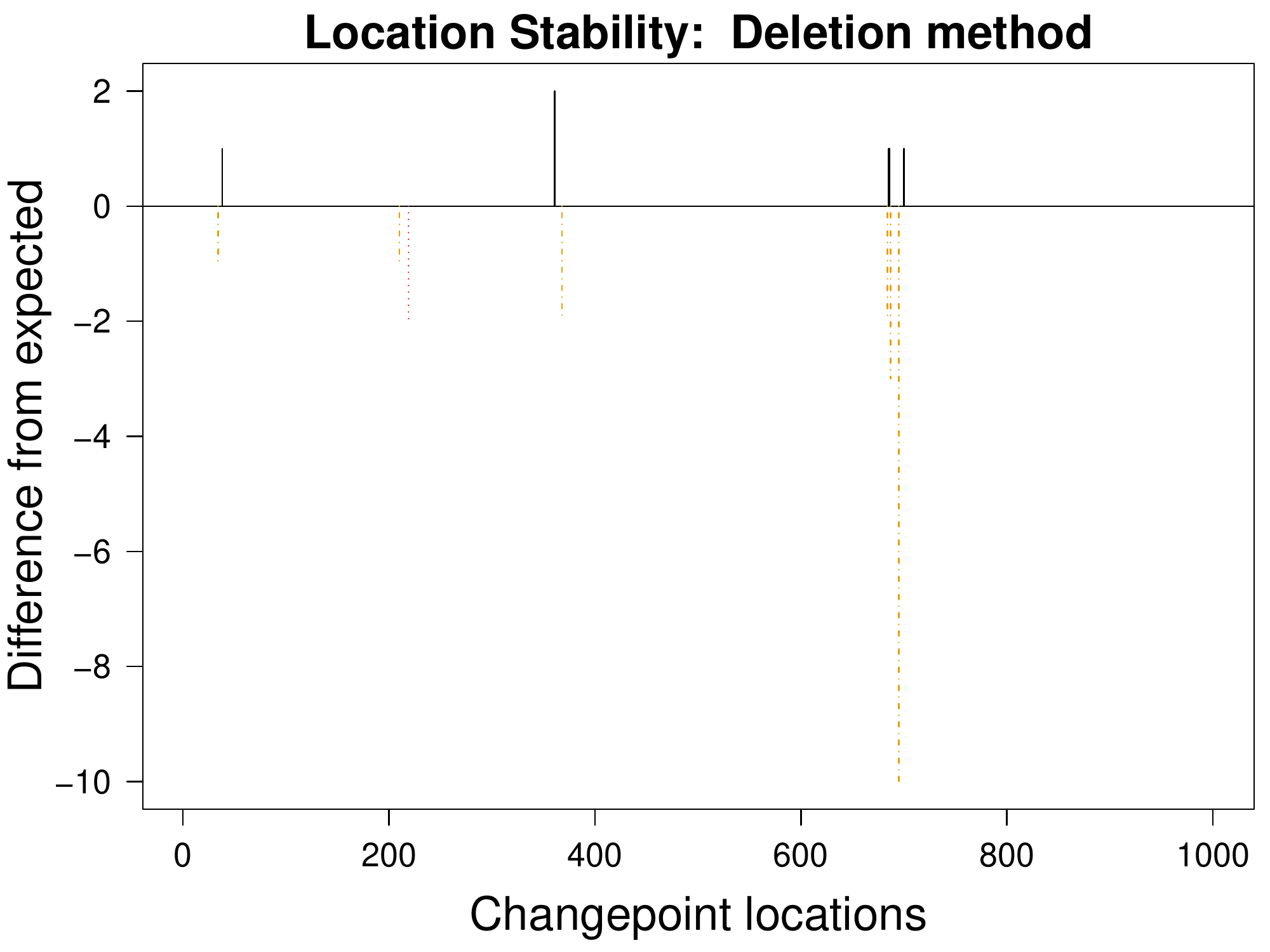}
\caption{\label{WellLogLocStab:del}}
\end{subfigure}
\begin{subfigure}[b]{0.49\textwidth}
\includegraphics[width=\textwidth, page = 2]{Figure10.pdf}
\caption{\label{WellLogLocStab:out}}
\end{subfigure}
\caption{\label{WellLogLocStab}Location Stability plots for the Well-log data.}
\end{figure}

Next, we address the question \textit{``How and where do the instabilities manifest?"}. First, consider the stability of the changepoint \textit{locations}, as visualized in the Location Stability plots of Figure \ref{WellLogLocStab}.
For the deletion method, 
\textcolor{black}{the few positive short (black) heights (Figure \ref{WellLogLocStab:del})}
immediately highlight that only a minority of location instabilities occur.
Hence, while many changes are labelled as potentially unstable (orange \textcolor{black}{dot-dashed lines in Figure \ref{WellLogLocStab:del}}); these instabilities only manifest themselves in rare cases.
For the \textcolor{black}{outlier} method, by contrast, especially changepoints 368 and 695 are prone to more severe instability 
\textcolor{black}{as can be observed from the long negative heights at their locations \textcolor{black}{in Figure \ref{WellLogLocStab:out}}.}
The  
\textcolor{black}{positive solid (black) heights} show that when the changepoints move, they do not move far from existing changepoint locations.
\textcolor{black}{The net negative balance shows that overall changepoints are deleted rather than moved.} \textcolor{black}{The analysis of the well-log data is, however, more complex than the simulation data example, which makes it harder to directly associate the changepoint moves (black positive lines) to the original changepoints (coloured negative lines) in the Location Stability plots. Practitioners are therefore advised to consult the more detailed Influence Map to match how various data perturbations affect the original data.}

\begin{figure}
\centering
\begin{subfigure}{0.49\textwidth}
\includegraphics[width=\textwidth, page = 1]{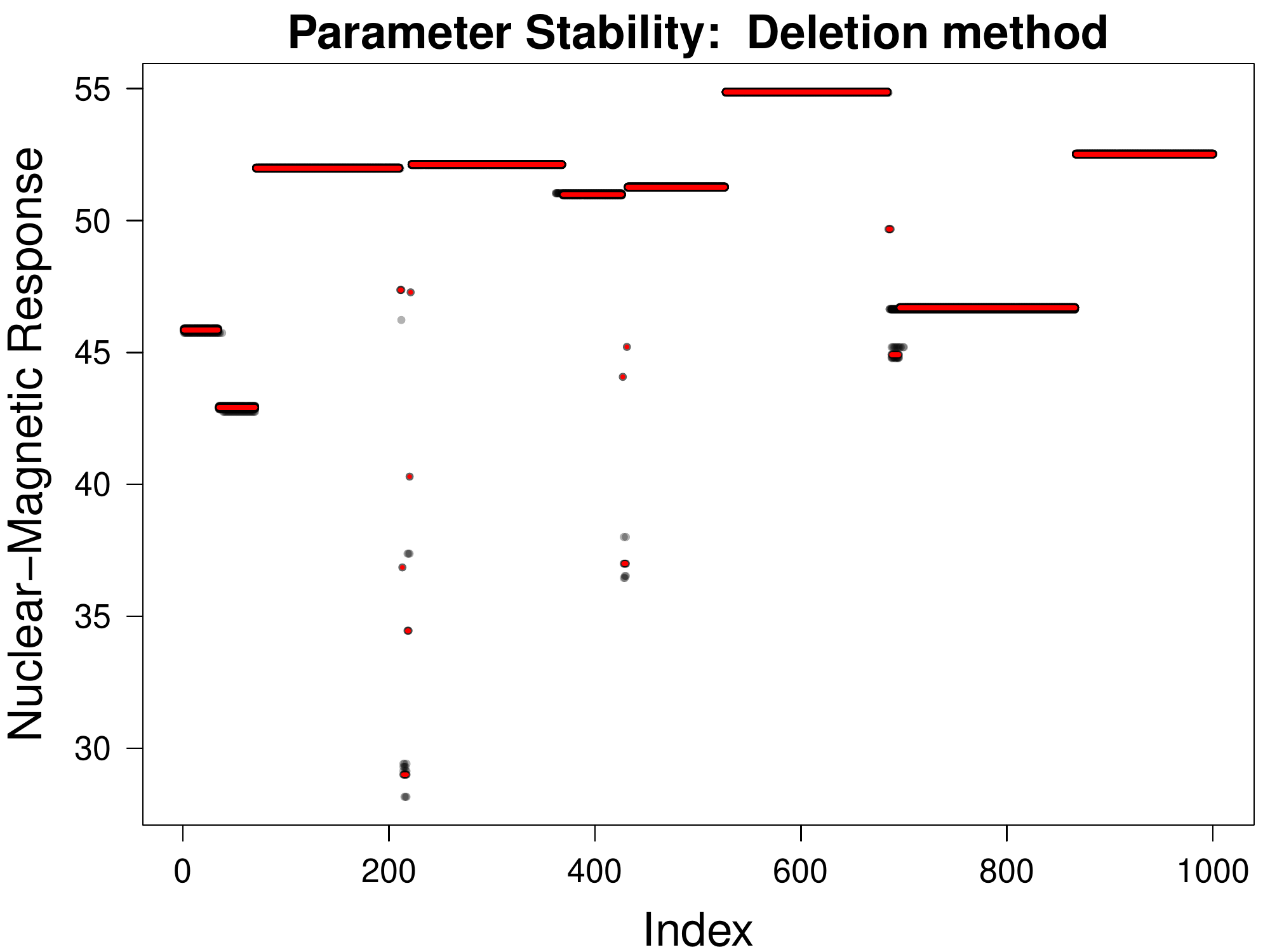}
 \caption{\label{fig:welllog:paramstab:del:detail}}
\end{subfigure}
\begin{subfigure}{0.49\textwidth}
\includegraphics[width=\textwidth, page = 2]{Figure11.pdf}
 \caption{\label{fig:welllog:paramstab:out:detail}}
\end{subfigure}
\caption{\label{WellLogParamStab}Parameter Stability plots for the Well-log data.}
\end{figure}

Secondly, consider the stability of the segment \textit{parameters}, namely the mean, as visualized in the Parameter Stability plots of Figure 
\ref{WellLogParamStab}.
Due to the (minor) evidence of changepoint location instability for the deletion method, the vast majority of segment means appears very stable in Figure \ref{fig:welllog:paramstab:del:detail}. Some instability can be observed for observations around changepoint 368 and 687 \textcolor{black}{by the additional dark lines occurring to their left. This indicates that these changepoints are (somewhat) unstable and cause their preceding observations to have lower segment means when the changepoint is moved earlier}.
These same instabilities in the segment mean appear even stronger in the Parameter Stability plot of the \textcolor{black}{outlier} method (Figure \ref{fig:welllog:paramstab:out:detail}).  
Furthermore, around observation 200, both parameter stability plots show some instabilities corresponding to the malfunctioning of the probe. When only two observations form a segment, the deletion or \textcolor{black}{contamination} of one of them causes the mean to be the other data point, thereby giving rise to the light gray areas of instability. 
Alongside this there are further instabilities in the 300-450 range.  Jointly using the Parameter Stability and the Location Stability plots we can see that this instability is driven by the instances where the change before 400 does not manifest along with the movement of the change after 400.  Anyone inspecting the original segmentation would be unlikely to think that a single data point would have such a profound and far-reaching effect on the resulting segment means.

\subsection{Sources of the Instabilities}
\begin{figure}
\centering
\begin{subfigure}{0.49\textwidth}
\includegraphics[width=\textwidth, page = 1]{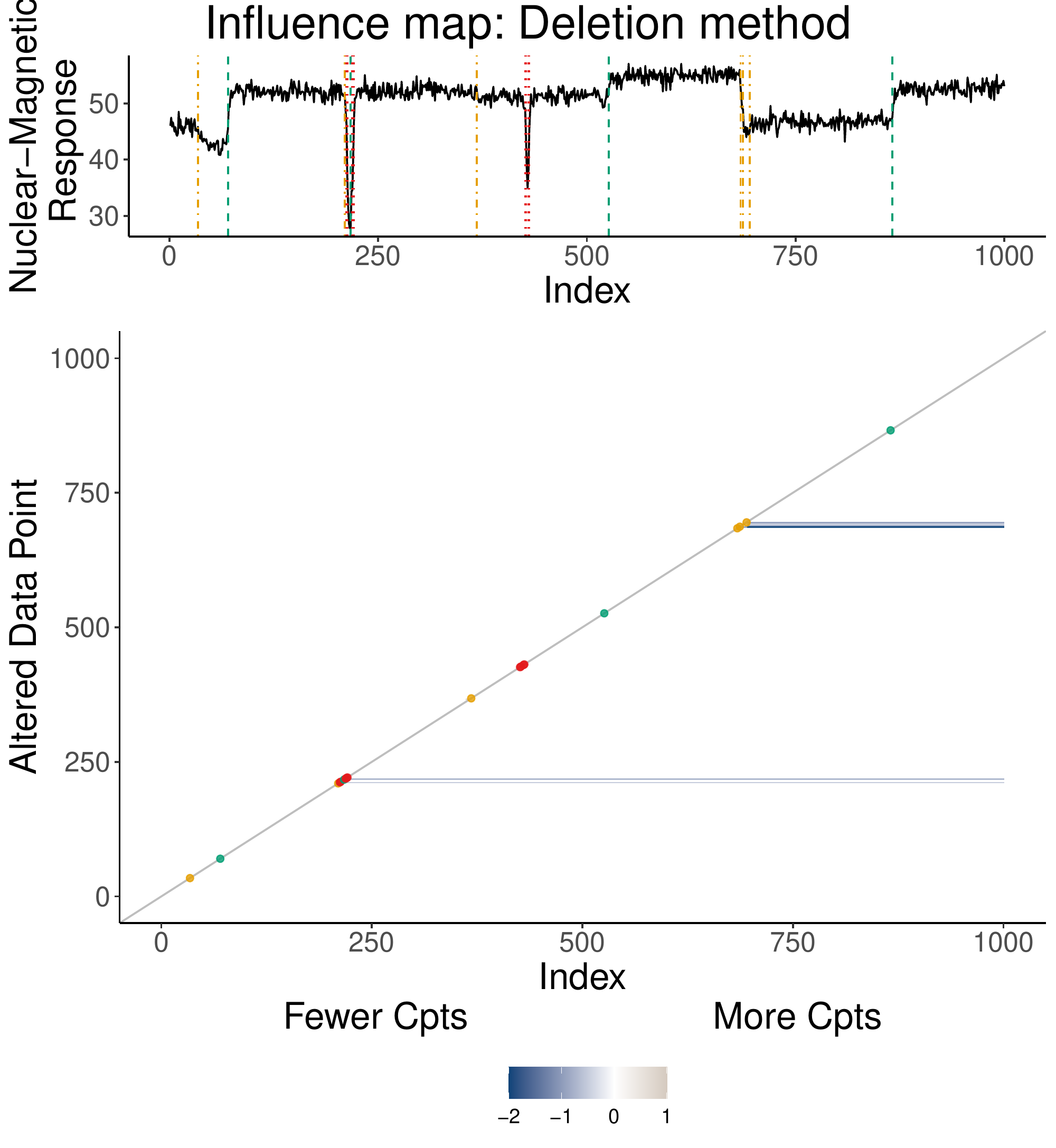}
 \caption{\label{fig:welllog:inflmap:del}}
\end{subfigure}
\begin{subfigure}{0.49\textwidth}
\includegraphics[width=\textwidth, page = 2]{Figure12.pdf}
 \caption{\label{fig:welllog:inflmap:out}}
\end{subfigure}
\caption{\label{WellLogInfMap}Influence Maps for the Well-log data.}
\end{figure}

Finally, we consider our more detailed influence Diagnostic Objective, namely \textit{``Which single influential observations trigger these instabilities to arise and how so?"}, through the lens of the Influence Maps (Figure \ref{WellLogInfMap}).

We first discuss the results for the deletion method (Figure \ref{fig:welllog:inflmap:del}). 
Several potentially unstable (orange) changepoints-- such as the one at location 34--hardly have any (clearly) visible coloured \textcolor{black}{pixels} of instability surrounding them. This is due to the larger dataset size $n=1000$ than our previous example.  We recommend to consider these changepoints as sufficiently stable. 
A handful of observations are found to assert notable influence.
Recall that within the data just after time points 200 and 400 there are malfunctions in the observations recorded.  Rather than affect single observations these result in a quick degradation and restoration of the signal.  Most, but not all, of these are isolated as individual segments in the original segmentation. The deletion of irregular data points 212, 218 and 219 each triggers the changepoint induced by their outlyingness to disappear.
Similarly, almost all observations in the range 685-698 are highly influential: their deletion triggers the segment 687-695 to no longer arise. 

Secondly, consider the results for the \textcolor{black}{outlier} method (Figure \ref{fig:welllog:inflmap:out}).
The largest area of influential observations (vertical axis) concerns the data points in the range 681-845.
Their \textcolor{black}{contamination} triggers changepoint 695 to disappear; the first five additionally trigger the removal of changepoints 684 and 687.  This is interesting as it signifies that if the original segment were shorter, the changepoint at 695 would no longer arise.  Care must be taken about inference made for this changepoint.

Interestingly, the same phenomenon occurs for the instability triggers 326-384:  their \textcolor{black}{contamination} causes changepoint 368 to disappear.
Other similar, though less outspoken, influential regions occur in Figure \ref{fig:welllog:inflmap:out}. All of these are blue, indicating a changepoint removal and the majority of them affect subsequent observations from the same segment (since the coloured \textcolor{black}{pixels} start below-diagonal and continue until the end of the sample). 

\section{Conclusion} \label{sec:conclusion}
Motivated by questions from practitioners in applying commonly used changepoint methods, this paper has presented the first approach to considering influence of the observed data points on changepoint segmentations.  We provide a framework for two methods to characterize influence; deletion and \textcolor{black}{contamination}. Alongside the framework three levels of graphics were introduced.  The stability dashboard provides an overview of the results which indicates if there are any locations of concern.  The location and parameter stability plots provide the second granularity of detail indicating how the segmentations are affected.  The most detailed level is the influence map which includes {\itshape which} observations are influential and {\itshape how} they influence the segmentation.

\textcolor{black}{A challenging aspect of the proposed approach is to characterize what a ``no problem" situation looks like.  The simplest answer is that if all changepoints are stable (dashed green) then there is no problem.  However, in reality this is unlikely to be the case as short segments, small changepoints and clustered changepoints as seen in our well-log example, are common.  We have deliberately not addressed this subjective issue of when a point is ``influential enough" to compromise an analysis, as the deliberation of this depends on the downstream pipeline of decision to be made based upon the original segmentation. We prefer to leave this evaluation to the sensibility of the practitioner.}

We illustrated our general approach using the change in Normal mean test statistic coupled with the PELT search method but we stress that our approach can be applied to {\it all} changepoint methods.  Furthermore, the only aspect of this paper which is specific to the change in Normal mean test statistic is our justification of the expected alterations to the segmentations that feed into the \textcolor{black}{location stability and} influence map.  For other test statistics these either need to be calculated or to plot the altered segmentations rather than the difference from the expected.  This is an important consideration as if one was using a robust test statistic such as that provided in \cite{Fearnhead19} then the outlier method would not guarantee the creation of two new changepoints.  It is still interesting to consider the influence of the data in this robust setting but we leave this for future work.  Our aim in this paper is to provide a framework for assessing influence in a general sense; utilising a common test statistic and search method purely as an example.  \textcolor{black}{In future research it would be interesting to explore whether different test statistic and search method combinations may be more/less prone to instabilities than others.}

Finally, one may consider that the influence plots characterize information about uncertainty in the changepoint segmentation.  Whilst this is true, we are not aiming to provide confidence intervals or similar measures of uncertainty quantification.  Akin to regression analyses, there are questions best answered by confidence intervals and others by measures of influence.  Similarly, we have advocated questions here that practitioners may wish to answer for which a measure of influence for changepoint segmentations is required.

\color{black}

\section*{Acknowledgments} 
Wilms gratefully acknowledges funding from  the European Union's Horizon 2020 research and innovation programme under the Marie Sk\l{}odowska-Curie grant agreement  No 832671.
Killick gratefully acknowledges funding from EP/R01860X/1, EP/T014105/1, NE/T012307/1 and NE/T006102/1. 
Matteson gratefully acknowledges funding from a Xerox PARC Faculty Research
Award, National Science Foundation Awards 1455172, 1934985, 1940124, and 1940276, USAID, and Cornell University Atkinson Center for a Sustainable Future.

\color{black}

\bibliographystyle{Chicago} 
\bibliography{refs}

\newpage

\setcounter{figure}{0}
\setcounter{table}{0}
\renewcommand{\thefigure}{A\arabic{figure}}
\renewcommand{\thetable}{A\arabic{table}}
\renewcommand{\theequation}{A\arabic{equation}}

\appendix 
\section*{Appendices}

\section{Background to Changepoint Methods Used}\label{sec:background}
The paper proposes a general influence framework for all changepoint problems.  However, the calculation of the expected segmentations under the 
data alterations
requires the assumption of an underlying model and inference framework.  We choose to demonstrate the framework on the simplest, but informative, changepoint model, the multiple changepoint problem where $y_1,\ldots, y_n$ is an $n$ length time series which is assumed to follow a Normal distribution with mean $\mu_{\{i\}}$ and variance $\sigma^2$.  The $\mu_{\{i\}}$ are assumed to follow a multiple changepoint structure with
\begin{align*}
\mu_{\{i\}}=\begin{cases}
        \mu_1 & \mbox{for } 0 < i \leq \tau_1 \\
        \mu_2 & \mbox{for } \tau_1 < i \leq \tau_2 \\
        \vdots & \vdots \\
        \mu_{m+1} & \mbox{for } \tau_m < i \leq \tau_{m+1},
       \end{cases}
\end{align*}
where the $\{\tau_i\}_{i=1}^m$ are the $m$ ordered changepoint locations and adopting the standard notational convention that $\tau_0=0$ and $\tau_{m+1}=n$.

We use a minimum penalized cost approach to infer the number of changepoints and their location. 
The cost associated with a segment of data ${\boldsymbol y}_{s:t}=(y_s, \ldots, y_t)$ is given by
$$ 
\mathcal{C}({\boldsymbol y}_{s:t}) = \underset{\theta}{\operatorname{min}} \sum_{j=s}^t \gamma(y_j; \theta),
$$
where $\gamma(y; \theta)$  is a loss function for a single observation $y$ and $\theta$ is a segment-specific location parameter.  In our simulations and data example, the loss function utilized is twice the negative log-likelihood for a Normal distribution (ignoring terms which are constant across segments),
$\gamma(y; \theta)= \frac{1}{\sigma^2}\left(y-\theta\right)^2$.

The penalized cost for a segmentation is then
\begin{align} 
\mathcal{Q}({\boldsymbol y}_{1:n}; \boldsymbol \tau_{1:m}) = \sum_{j=0}^m \left( \mathcal{C}({\boldsymbol y}_{\tau_{j}+1:\tau_{j+1}})  + \beta \right), \label{eqn:tooptimize}
\end{align}
where $\beta>0$ is a penalty cost for the introduction of a changepoint and the vector $\boldsymbol \tau_{1:m}$ collects all changepoint locations. 

In order to estimate the number and location of the changepoints, we need to minimize problem \eqref{eqn:tooptimize}.  There are several approaches which, directly or indirectly, optimize a form of \eqref{eqn:tooptimize}.  We choose to use the PELT algorithm \citep{killick2012optimal} as it is an exact optimizer of problem \eqref{eqn:tooptimize}.  PELT uses dynamic programming to rearrange \eqref{eqn:tooptimize} to
\begin{equation}
\mathcal{Q}({\boldsymbol y}_{1:n}; \boldsymbol \tau_{1:m}) = \min_{j=1:n}\left( \mathcal{Q}({\boldsymbol y}_{1:\tau_j}) + \mathcal{C}({\boldsymbol y}_{\tau_{j}+1:n})  + \beta 
\right), \label{eqn:OP}
\end{equation}
where $\mathcal{Q}({\boldsymbol y}_{1:\tau_j})$ is the cost of the optimal segmentation for data $y_1,\ldots,y_{\tau_j}$.  The final solution in equation \eqref{eqn:OP} is not calculated directly but is calculated for increasing lengths of the data $y$.  Provided that the loss function for a portion of data reduces when a changepoint is added (i.e.\ more parameters improves the fit), PELT further prunes the minimization over $j$ to reduce computational time.  See \cite{killick2012optimal} for full details.

\color{black}

\section{Results on the Expectation} \label{sec:theory}
We provide numerical and theoretical results on the segmentation one expects under the two data alternations: deletion and \textcolor{black}{contamination}.
We assume the inference is conducted using a penalized cost approach as discussed in Appendix \ref{sec:background} with the squared error loss 
$
\gamma(y; \theta) = (y - \theta)^2
$
which is commonly used for inferring the mean of the data, but other choices can be made.
\textcolor{black}{Note that this is a scaled version of our negative log-likelihood loss.}

\subsection{Deleting Observations} \label{loo}
We investigate how one should expect a deletion of a single data point to affect the changepoint segmentation.
We generate ordered sequences of length $n=\{100, 200,$ $300, 400, 500, 1000\}$  with one change in mean: observations belonging to the first half of the sample are generated from a $N(0, 1)$; observations belonging to the second half of the sample from a $N(\delta, 1)$, where we consider different values for the size of the change $\delta=\{1, 2, 3, 4, 5\}$ and the variance is kept fixed at one.  

We then apply our rolling procedure thereby subsequently deleting each observation $t=1, \ldots, n$ and re-estimating the changepoint location with PELT.
For each observation $t=1, \ldots, n$ in each simulation run, we record whether the changepoint moves beyond its original location. We set the number of repetitions of each simulation scenario to $500$.

Note that our expectation is that observations deleted prior to the true changepoint location will see the estimated changepoint location reduce by one whereas observations deleted after the true changepoint location will remain the same. 

\begin{figure}[t]
\centering
\includegraphics[width=0.5\textwidth]{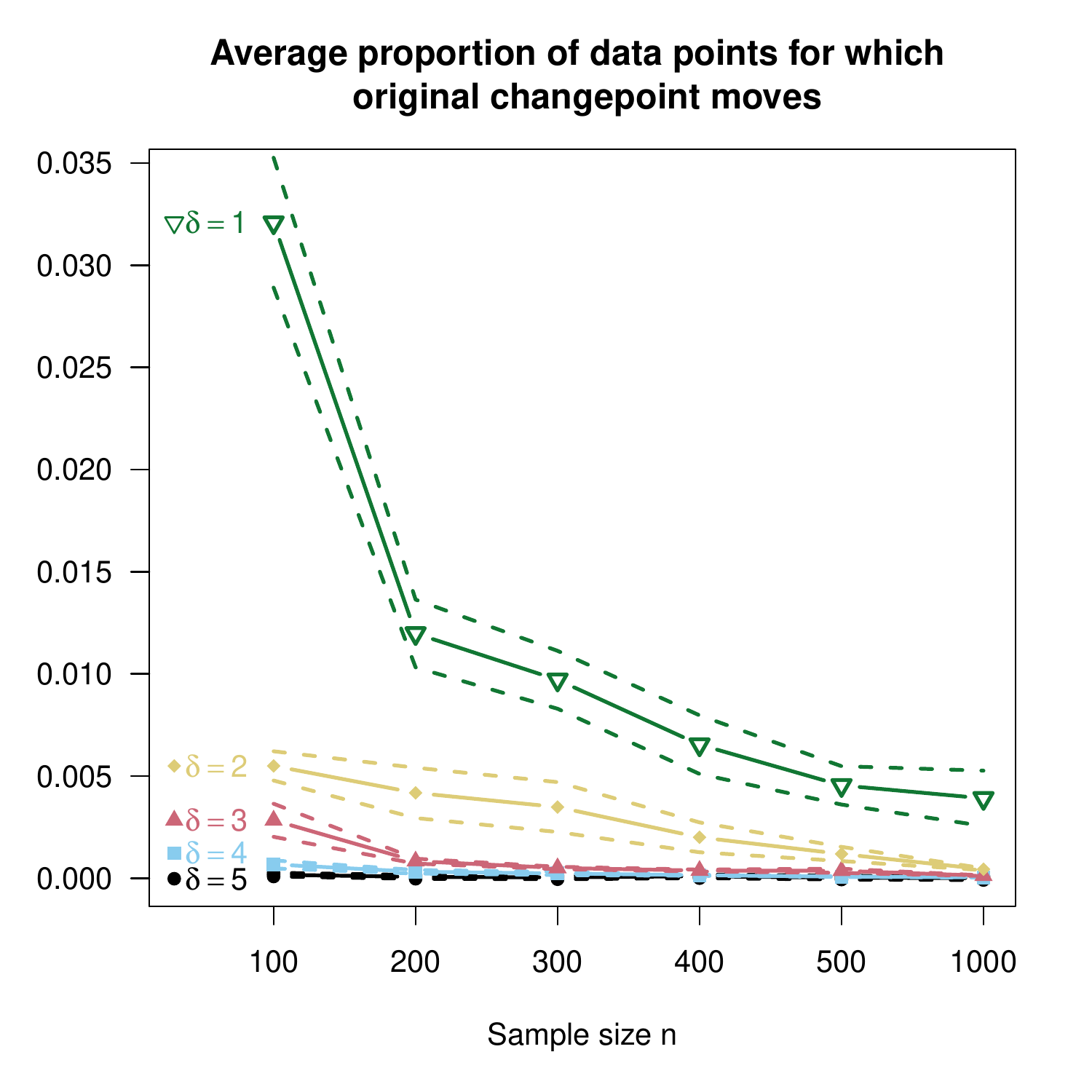}
\caption{Simulation results for Deletion method: Average proportion of data points for which the original changepoint moves, for different values of the shift size $\delta$ and sample size $n$.
Intervals of length two standard errors are displayed by the dashed lines.
\label{simulation_plots}}
\end{figure}

In Figure \ref{simulation_plots}, we plot the average (taken over all simulation runs) proportion of data points for which the original changepoint moves, and this for different values of the shift size $\delta$ and sample size $n$.
\textcolor{black}{
Intervals of length two standard errors are indicated by the dashed lines.}
As with all changepoint problems, the asymptotics depend on the size of the change and the segment length.
As the sample size $n$ and/or the shift size $\delta$ increases, we observe-- as expected--almost no changepoint moves beyond its original location. 
For small mean shifts and/or sample sizes, minor perturbations from the expectation are detected. 
These exactly correspond to changepoint instabilities practitioners should be warned for and will thus show up in the Influence Map.

\subsection{\textcolor{black}{Contaminating} Observations} \label{moo}
We numerically investigate how one should expect a contamination of a single data point, namely when made outlying, to affect the changepoint segmentation. 
Theorem 1 in \cite{Fearnhead19} shows that for a sufficiently large outlier, $y_t$,  two additional changepoints at both $t-1$ and $t$ reduce the penalized cost. The optimal segmentation is thus expected to have two additional changes at these locations. The next proposition extends Theorem 1 in \cite{Fearnhead19}  to prove how large the outlier needs to be in order for it to introduce two additional changes. 

\begin{proposition} \label{prop_outlier}
Assume that the loss function satisfies $\gamma(y; \theta) = g(|y - \theta|)$, where $g(0) = 0$, and $g(\cdot)$ is an unbounded, increasing function. Choose any $t \in \{1, \ldots, n\}$ and fix the set of other observations $y_s$ for $s\neq t$.
Let $\hat{\theta}_{s:u}$ be the minimizer of the cost function on the current segment containing time point $t$.
Then for values of  $y_t$ satisfying $2\beta < \gamma(y_t; \hat{\theta}_{s:u})$ the segmentation that minimizes the penalized cost will have changepoints at $t-1$ and $t$.
\end{proposition}
\begin{proof}
We follow the proof of Theorem 1 in \cite{Fearnhead19} to show how large the value of an atypical data point $y_t$ needs to be for the optimal segmentation to have changes at both $t-1$ and $t$ (or only one of these if the original segmentation already has a change at the other time-point).

Consider any segmentation of the data that does not include changepoints at both $t-1$ and $t$. 
Let the segment of the original segmentation that contains the outlier $y_t$ be $\boldsymbol y_{s:u}$ for $s < t$ and $u > t$. The change in cost between this segmentation and the segmentation with additional changepoints at $t-1$ and $t$ is
\begin{equation}
\operatorname{min} \sum_{j=s}^{t-1} \gamma(y_j; \theta) + \underset{\theta}{\operatorname{min}} \sum_{j=t+1}^{u} \gamma(y_j; \theta) + 2\beta - \operatorname{min} \sum_{j=s}^{u} \gamma(y_j; \theta), \label{changecost}
\end{equation}
see  \cite{Fearnhead19}.

If the change in cost in \eqref{changecost} is positive, no additional changepoints will be induced.
If the change in cost in \eqref{changecost} is negative, two additional changepoints will be induced. 
We will show how large $y_t$ needs to be for the latter to occur.

For convenience, we first introduce the following notation:
$$
\hat{\theta}_{s:(t-1)} = \underset{\theta}{\operatorname{argmin}} \sum_{j=s}^{t-1} \gamma(y_j; \theta), \ \ 
\hat{\theta}_{(t+1):u} = \underset{\theta}{\operatorname{argmin}} \sum_{j=t+1}^{u} \gamma(y_j; \theta), \ \text{and} \ 
\hat{\theta}_{s:u} = \underset{\theta}{\operatorname{argmin}} \sum_{j=s}^{u} \gamma(y_j; \theta).
$$
Under a positive change in cost, \eqref{changecost} becomes
\begin{equation}
\sum_{j=s}^{t-1}\gamma(y_j; \hat{\theta}_{s:(t-1)}) + \sum_{j=t+1}^{u} \gamma(y_j; \hat{\theta}_{(t+1):u}) + 2\beta  \geq \sum_{j=s}^{u} \gamma(y_j; \hat{\theta}_{s:u}) \nonumber \\
\label{changecosthat}
\end{equation}
where 
$$
\sum_{j=s}^{u} \gamma(y_j; \hat{\theta}_{s:u}) =  \sum_{j=s}^{t-1}\gamma(y_j; \hat{\theta}_{s:u}) + \gamma(y_t; \hat{\theta}_{s:u}) +  \sum_{j=t+1}^{u} \gamma(y_j; \hat{\theta}_{s:u}).
$$
Using the fact that for likelihoods (without penalization) adding a changepoint is always preferred (lower cost) we have that 
no additional changes will be induced if
$2\beta \geq \gamma(y_t; \hat{\theta}_{s:u})$. Two additional changes will be induced if 
$2\beta < \gamma(y_t; \hat{\theta}_{s:u})$.
\end{proof}

The proposition shows that 
two additional changes occur if the cost $\gamma(y_t; \hat{\theta}_{s:u})$ of keeping the observation in the current segment is larger than the cost $2\beta$ of introducing two additional changes.
Numerical simulations (unreported) confirm this result.

Since the values of $y_t$ resulting in two additional changepoints depend on the original data at hand, it would be computationally cumbersome to calculate the exact boundary for each \textcolor{black}{contaminated} point.  
Thus, to avoid computational overload in computing the cost of each data point, we set the value of the \textcolor{black}{contaminated} data point equal to twice the data range to construct our influence diagnostic plots in practice. \textcolor{black}{Simulation experiments (unreported) confirmed that \textcolor{black}{contaminating} data points by adding twice the range of the data is sufficiently large to guarantee the expected number of additional changes across different considered sample sizes.}

\end{document}